\documentclass[12pt]{article}
\usepackage{graphicx,amsmath,amssymb,cite}
\usepackage{epsfig,epsf}
\usepackage{graphicx}
\usepackage{epstopdf}
\newcommand{\beq}{\begin{equation}}
\newcommand{\eeq}{\end{equation}}
\newcommand{\alphabar}{\overline{\alpha}_s}
\newcommand{\betabar}{\bar{\beta_0}}

\numberwithin{equation}{section}

\renewcommand{\thefootnote}{\fnsymbol{footnote}}

\begin{document}

\begin{flushright}
DESY 17-090\\
\end{flushright}
\vspace*{0.5 cm}

\begin{center}


{\Large{\bf Decoupling of the leading contribution  in the discrete BFKL Analysis \\ of High-Precision HERA Data.  }}

\vspace*{1 cm}

{\large H. Kowalski~$^1$, L.N. Lipatov~$^{2}$\footnote[2]{Died September 4, 2017}, D.A. Ross~$^3$, O. Schulz~$^4$ } \\ [0.5cm]
{\it $^1$ Deutsches Elektronen-Synchrotron DESY, D-22607 Hamburg, Germany}\\[0.1cm]
{\it $^2$  St.Petersburg State University, St. Petersburg 199034 \\
 and  Petersburg Nuclear Physics Institute, Gatchina 188300,  Russia
}\\[0.1cm]
{\it $^3$ School of Physics and Astronomy, University of Southampton,\\Highfield, Southampton SO17 1BJ, UK}\\[0.1cm]
{\it $^4$ Max Planck Institute for Physics, Munich,  Germany}\\[0.1cm]
 \end{center}

\vspace*{3 cm}

\begin{center}
{\bf Abstract}  \end{center}
We analyse,  in NLO, the  physical properties of the discrete eigenvalue solution for the BFKL equation. We show that a set of
 eigenfunctions with positive eigenvalues, 
 $ \omega$, together with a small contribution from a continuum of eigenfunctions with
 negative $ \omega$, provide an excellent description of high-precision HERA $F_2$ data in the region, $x<0.001$, 
$Q^2 > 6 $ GeV$^2$. The phases of the eigenfunctions can be obtained from a simple parametrisation of the pomeron spectrum, which has a natural motivation within BFKL. The data analysis shows that  the first eigenfunction  decouples completely or almost completely from the proton. This  suggests that there exist an additional ground state, 
which is naturally saturated and may have the properties of the soft pomeron.  



\vspace*{2.5 cm}

\begin{flushleft}
    Oct  2017   \\
\end{flushleft}

\newpage 

\renewcommand{\thefootnote}{\arabic{footnote}}

\section{Introduction}

\begin{figure}[htbp]
\centerline{\includegraphics[width=10cm,angle=0]{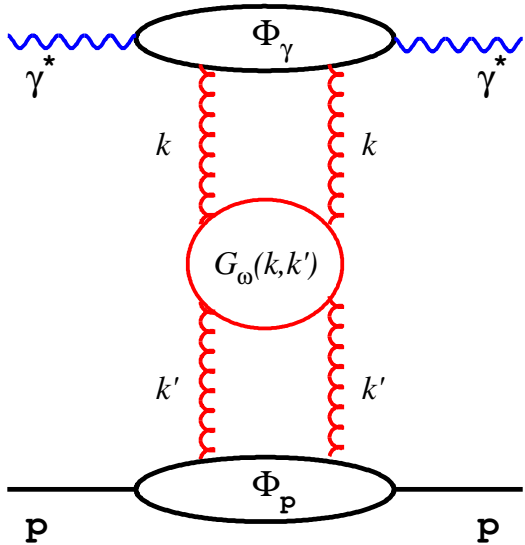}}
\caption{ Evaluation of $F_2$ in $\gamma^*$ p scattering using   the BFKL Green Function    }
\label{gpin} \end{figure}

The  aim of this paper is to apply,  for the first time, the complex BFKL Green Function  approach developed in  our two previous papers~\cite{LKR1,KLR16}  to the analysis of HERA data,.  The new approach, although  seemingly equivalent to the discrete BFKL solution developed in our previous
 papers~\cite{KLRW,KLR}, exhibits some differences. The most important of which is that the normalisation of the eigenfunctions is now determined analytically instead of being  determined only numerically, as was the case in ref.~\cite{KLRW,KLR}.  As shown in~\cite{KLR16},  this seemingly minor technical difference
 has  an important consequence: the convergence of the eigenfunctions is now much more rapid than previously. Instead of using O(100) eigenfunctions, as in  
  ref.~\cite{KLRW,KLR}, we need  to use  only O(10) eigenfunctions to properly represent  the Green Function. This constrains substantially   the new BFKL solution, exhibits more clearly its physical properties, and leads to new results.
 
 To obtain a good description of the HERA $F_2$ data  it is necessary to define a non-perturbative boundary condition defined in terms of phases of eigenfunctions at low gluon transverse momenta $k$, close to $k \sim \Lambda_{QCD}$. Since in~\cite{KLRW,KLR} we were using a large number of eigenfunctions, O(100), it was easy to find a simple, ad hoc, parametrisation for these phases. However, this parametrisation had no physical interpretation.  
 
 The first task of this paper is to find a simple parametrisation for the phases of much fewer eigenfunctions, O(10). 
  In the search for such a condition we are guided by the principle of simplicity and some analogy to the Balmer series. 
  In the QCD version of  Regge theory developed in our papers, the BFKL  equation is considered to be analogous to the Schr\"odinger equation for the wavefunction of the pomeron.  The BFKL kernel corresponds to the Hamiltonian and the eigenvalues $\omega$  to the  energy  eigenvalues. 
 In this paper, we find  that we can specify the boundary condition in terms of a relation between the eigenvalues $\omega_n$ of the BFKL operator and the principal quantum number $n$. This relation then determines the boundary condition in terms of the  phases $\eta_n$ of the eigenfunctions,  close to the non-perturbative region,  $k \sim \Lambda_{\rm QCD}$. In addition, the relation between  $\omega$  and $n$ is very simple and, for large $n$,
has  a good physical motivation within the context of the  BFKL formalism.
 
We show in this paper that this new approach leads to unexpected results and gives a new insight into the role of gluon density. We recall that
 the BFKL Green Function  is directly related to the gluon density (see below). The properties of this gluon density  are very interesting for
 the LHC and cosmic ray physics. They are also interesting in themselves, because in contrast to the DGLAP evolution~\cite{DGLAP},  the BFKL equation  describes a system of quasi-bound self-interacting gluons. Such a system is sensitive to confinement effects and  also has  some  sensitivity to Super-Symmetry effects (in the gluon sector), as was first observed in ref.~\cite{KLRW,KLR} and is also valid in the present approach.



 The paper is organised as follows:  In Section 2 we recall the main properties of the BFKL Green Function and of their  eigenfunctions, determined in our last papers~\cite{LKR1,KLR16}. We also indicate here the differences between the approach of ref.~\cite{KLRW,KLR}  and our present approach. In Section 3 we introduce the NLO corrections to BFKL and evaluate the properties of eigenvalues and eigenfunctions at NLO. In Section 4 we apply this formalism to HERA data and describe the search for a proper boundary condition and the new  results.
 Finally, in Section 5 we summarise the results and conclude.

\section{BFKL Green Function } 
\label{sec0}

 The Green Function approach considered here  is highly appropriate since it does not require any cutoff on the BFKL dynamics and provides a direct relation to the measurements at low-$x$.    Thus, the deep inelastic  structure function $F_2(x,Q^2)$  can be directly calculated as a  convolution
of the Green function with  impact factors that encode the coupling of the
 Green function to the external particles that participate in that process. 
 \beq F_2(x,Q^2)  \ = \ \int dt dt^\prime \Phi_{\gamma}(Q^2,t) {\cal G}(t,t^\prime,Y)
  \Phi_P(t^\prime), \eeq
where, $Y=\ln(1/x)$, $t=\ln(k^2/\Lambda_{QCD}^2), \ t^\prime=\ln(k^{\prime \,2}/\Lambda_{QCD}^2)$; 
$k, \ k^\prime$ being the transverse momenta of the gluons entering the
 BFKL  amplitude.
 $\Phi_{DIS}(Q^2,t)$ describes the (perturbativly calculable) coupling of the gluon with transverse momentum $k$
 to a photon of virtuality $Q^2$ and $\Phi_P(t^\prime)$ describes
the coupling of a gluon of transverse momentum $k^\prime$ to the target proton, see Fig.\ref{gpin}. \footnote{The variable $t$ is more appropriate for theoretical analysis, whereas $k$ is more appropriate for comparison with data. To translate $t$ to $k$ we assumed that  $ \Lambda_{QCD} = 275$ MeV.}

In ~\cite{LKR1} we determined  the BFKL Green Function ${\cal G}_\omega(t,t^\prime)$ 
(in Mellin space)  from the  equation
\beq \left( \omega- \hat{\Omega}(t,\hat{\nu}) \right) {\cal G}_\omega(t,t^\prime)
 \ = \ \delta(t-t^\prime), \label{BFKLeq} \eeq
where $\hat{\Omega}$ denotes the  BFKL operator, which was given in terms of the 
LO characteristic function, $\chi(\alpha_s(t),\nu)$, by
\beq \hat{\Omega} \ = \   \sqrt{\alphabar(t)} \left( 
 2\Psi(1) - \Psi\left(\frac{1}{2}+\frac{\partial}{\partial t} \right)
-\Psi\left(\frac{1}{2}-\frac{\partial}{\partial t} \right) 
 \right)  \sqrt{\alphabar(t)}, \label{eqomop} \eeq
 with  
$ \alphabar \, \equiv \, C_A \alpha_s/\pi$.
 By placing $\sqrt{\alphabar(t)}$ on either side
of the differential operator we assured the hermiticity of the whole operator. 

We have shown in~\cite{LKR1,KLR16}  that the Green Function determined in this way has poles on the positive real axis of the $\omega$ plane and a cut along the negative   
$\omega$  axis. Therefore it can be constructed from the complete set  of eigenfunctions of the  BFKL operator in the usual way
  \beq {\cal G}(t,t^\prime,Y)  \   = \ \sum_{n=1}^\infty x^{-\omega_n} f_{\omega_n}((t) f^*_{\omega_n}(t^\prime)
 +  \lim_{\omega_{min}\to -\infty} \int_{\omega_{min}}^0
  d\omega x^{-\omega} 
     f_{-|\omega|}(t) f_{-|\omega|}(t^\prime)
   .   \label{greenf}\eeq
The spectrum  of the eigenvalues $\omega_n$ was found to be discrete for positive values of $\omega$ and continuous for negative value of  $\omega$.
The complete set of eigenfunctions with positive and negative eigenvalues $\omega$  was found to satisfy  the closure relation and the orthonormality condition. In addition,   the Green Function  converges rapidly so it was sufficient to use only O(10)  discrete eigenfunctions
(see the discussion below eq.(\ref{norm1})) to describe properly the gluon density, as compared to our previous work~\cite{KLRW,KLR}, where we needed more than 100 eigenfunctions.


\subsection{Eigenvalues and eigenfunctions } 
\label{sec1}

 
 In LO BFKL~\cite{BFKL}, with  fixed QCD coupling constant $\alpha_S$,
 the eigenfunctions  have a simple oscillatory behaviour in terms of the gluon transverse variable $t$,
\beq f_{\omega}(k) \ \sim \ \exp(\pm i \nu t),  \label{fixeigen} \eeq
  The frequency $\nu$ of these oscillations is connected to the eigenvalue $\omega $  by the characteristic equation    
\beq \omega \ =   \overline{\alpha}_S \, \chi_0 ( \nu),  \label{fixval} \eeq 
with
  \beq \chi_0(\nu) \ \equiv \ 2\Psi(1) - \Psi\left(\frac{1}{2}+i\nu\right) 
 - \Psi\left(\frac{1}{2}-i\nu\right).  \eeq
With fixed  $\overline{\alpha}_S$  the frequency $\nu$ is a one-to-one function of $\omega$. However, when  $\alpha_S$ is running $\nu$  becomes a function of $t$, $\nu_\omega (t)$,  in order to compensate the $t$ variation of   $\overline{\alpha}_S$.  
 For sufficiently large values of $t$ there is no  real solution for $\nu_\omega(t) $ of eq.(\ref{fixval}). The transition from the real to imaginary values of $\nu_\omega(t)$ singles out a special value of $t=t_c$ for which
\beq  \nu_\omega (t_c) = 0.  \label{tcrit} \eeq 
For    values of $t$  below the critical point $t_c$ the behaviour of  the eigenfunction remains oscillatory, but above it becomes exponentially attenuated. This fixes the phase of the eigenfunction at $t=t_c$ and together with some fixed non-perturbative phase $\eta_{np}$ leads to quantisation, i.e to a discrete set of eigenfunctions.

To analyse the behaviour of the BFKL equation in the neighbourhood of the turning point,  $t_c$,
 it is convenient to define first  two related variables, 
$s_\omega (t)$ and $z(t)$.
 The variable $s_\omega (t)$ gives the phase shift from the turning point $t_c$ to the point $t$  and corresponds to the argument of the wave function of eq.(\ref{fixeigen}). It
 is defined as   
\beq
s_\omega (t) \ = \  \int _{t}^{t_c} dt'\,\nu _\omega (t')\,   \label{action}
\eeq
and the ($\omega$ dependent) variable $z(t)$ is defined as 
\beq
z (t)=   -  \left( \frac{3}{2} s_\omega (t) \right)^{\frac{2}{3}} .  \label{zvar}
\eeq
Using these variables we have shown in~\cite{LKR1} that 
 the BFKL  operator, $\hat{\Omega}$, 
 can be  related  to  the ``generalized Airy operator'' as 
\beq \left(\omega - \hat{\Omega}\left(t, -i\frac{\partial}{\partial t} \right)\right)
 \ 
\approx \ \frac{1}{N_\omega(t)} \left(\dot{z} z 
- \frac{\partial}{\partial t} \frac{1}{\dot{z}} 
 \frac{\partial}{\partial t} \right)  \frac{1}{N_\omega(t)} \label{eqi}. \eeq
 In this derivation the diffusion approximation was used in the vicinity of the turning point and the semi-classical approximation  far away from it. 
Using these approximations   we have shown~\cite{KLR16,LKR1} that the most general solution to equation~\ref{eqi}  is given by the Green Function 
\beq {\cal G}_\omega(t,t^\prime) \ = \ \pi 
  N_\omega(t)N_\omega(t^\prime)
 \left[ Ai(z(t)) \overline{Bi}(z(t^\prime) \theta(t-t^\prime) + t \leftrightarrow t^\prime
 \right]   \label{sol1a} ,\eeq
 with 
 \beq \overline{Bi}(z(t)) \ = \ Bi(z(t))
 + \cot\left(\phi(\omega)\right)Ai(z(t)). 
 \label{transform} \eeq
Here $Ai(z)$ and  $ Bi(z)$ denote the two independent  Airy functions.
The  function  $\phi(\omega)$ is defined as 
 \beq \phi(\omega) \ = \ s_\omega(t_0) + \frac{\pi}{4} -  \eta_{np}(\omega , t_0)\,  \label{phicon} \eeq 
with $\eta_{np}(\omega)  $  being a non-perturbative phase, fixed at some small $t_0$.   From~\ref{sol1a} and~\ref{transform} it follows, as discussed in ref.~\cite{LKR1},      that the BFKL Green function has poles when
\beq 
 \phi(\omega)=n\pi,  \ \  n= \ 0, \ Ê1, \ 2, \  3 \  ...\   . \label{quacon} 
\eeq 
The equations~\ref{quacon} and~\ref{phicon} define the eigenvalues   $\omega_n$, which are a function of the non-perturbative boundary condition $\eta_{np}(n)$.

Furthermore,  in~\cite{LKR1} we have shown that,
in case of  positive $\omega_n$,
the eigenfunctions of the BFKL
 operator  are given by
\beq f_{\omega_n}(t) \ = \ \sqrt{\frac{\pi}{\phi^\prime(\omega_n)}} N_{\omega_n}(t)Ai(z(t)),  \label{normeigen} \eeq 
whith  $N_{\omega_n}(t)$ being the normalisation factor, which is given by
 \beq N_{\omega_n}(t) \ = \
 \frac{|z(t)|^{1/4}}{\sqrt{\frac{1}{2} \alphabar(t) \chi^\prime\left(\nu_{\omega_n}(t) \right)}}\  .
\label{norm1} \eeq 
Here $\chi$ denotes the BFKL characteristic function which in LO is simply equal to $\chi_0$ but is more complicated in  NLO. 

The above expression is similar to the eigenfunctions used in ref.~\cite{KLRW,KLR} with the difference that  the normalisation factor, $N_{\omega_n}$, was not $t$ dependent  and  was determined by numerical integration.   In the first paper, in which we developed our new approach~\cite{LKR1}, we argued that this difference is not very important because the $t$ dependence of the normalisation factor is very slow and would not sizeably change the shape of the eigenfunctions.
Whereas  this is correct for the shapes  in the physical region, it   is
 {\it not} true for the normalisation.  The numerical integration, which determines the normalisation  factor, extends to very large $t$ regions (given by $t_c$, see Fig.~\ref{tc}),  much above the physical region. Therefore,
  enhanced $t$ dependence in~\cite{KLRW,KLR}  has a substantial effect when integrated over  large $t$ regions.  As explained  in~\cite{KLR16} the eigenfunction of eq.(\ref{normeigen}) converge as $1/n^2$, whereas these of ref.~\cite{KLRW,KLR}  converge on a much slower pace,  as $1/n$.

 To understand the physical meaning of the function $\phi$ it is useful to asymptotically expand   the Airy function  of eq.(\ref{normeigen}),  around $t=t_0$ (but far away from  $t_c$),
\beq
f_{\omega_n}(t_0) \ \propto \  Ai(z(t_0)) \ \approx \ \frac{1}{\sqrt{\pi} \ |z(t)|^{1/4}} \sin \left( s_\omega(t_0) + \frac{\pi } { 4} \right) \label{sexp} .
\eeq  
This means that
the function $\phi$ is the difference between 
 the perturbative and non-perturbative phases of the wave function, which should not depend on  $t_0$.~\footnote{Although we call this phase non-perturbative we fix it in the perturbative region, at $t_0$ equivalent to $k_0 = 1 $ GeV, close to $\Lambda_{QCD}$. At this $k_0$ the value of  $\alphabar $  is 0.50. }

For negative values of $\omega$ eq.(\ref{tcrit}) has no solution, i.e. there is  no critical point and no quantization of eigenvalues. The negative $\omega$ eigenfunctions were derived in~\cite{KLR16} and are given by 
\beq f_{-|\omega|}(t) \ = \ \sqrt{\frac{2}{\pi}} \frac{1}{\sqrt{\alphabar(t) \chi^\prime\left(\nu_\omega(t)\right)}}
 \sin\left( \int_{t_0}^t \nu_\omega(t^\prime) dt^\prime + \eta_{np}  \right). \label{negomega}  \eeq

The eigenfunctions defined by eq.(\ref{normeigen})  and (\ref{negomega})   fulfil the completeness relation 
\beq \lim_{\omega_{min} \to -\infty} \int_{\omega_{min}}^0 d\omega f_{-|\omega|}(t) f_{-|\omega|}^*(t^\prime) 
 +  \sum_{n=1}^\infty f_{\omega_n}(t)f_{\omega_n}(t^\prime)   \ = \ \delta(t-t^\prime) \label{complete}  \eeq
and are orthonormal, as shown in~\cite{KLR16}.

\section{NLO evaluation}

To obtain the eigenfunctions of the BFKL equation in NLO we  just need 
to replace  eq.(\ref{fixval}) by its NLO counterpart
\beq \omega \ = \ \alphabar \chi_0(\nu) + \alphabar^2 \chi_1(\nu)
 + {\cal O}(\alphabar^3) \label{pert13} \eeq
where $\chi_0(\nu)$ and $\chi_1(\nu)$ are the LO and NLO characteristic functions
 respectively. The NLO value of $\alpha_s$ was fixed by measurement at  $Z^0$ pole. In our numerical analysis, we modify $\chi_1$ following the method of Salam \cite{salam} in which the collinear contributions are resummed, leaving a remnant which is accessible to a perturbative analysis. For the analysis of this paper we use Scheme 3 of ref.~\cite{salam} (see Appendix A).
 
To create the eigenfunctions we have chosen  the value of $t_0$ equivalent to $k_0 = 1 $ GeV, close to $\Lambda_{QCD}$ but still in the perturbative region, 
  with  $\alphabar(k_0) = 0.50 $. To be able to describe the measured structure function $F_2$, which has a changing  slope $\lambda$,  $\eta_{np}$  should vary with $n$ and  
  the value of the non-perturbative phase $\eta_{np}$ for the leading eigenfunctions  should be close to zero  (see the discussion  in  Sections 4.1 and 4.3). 
   We  have therefore adopted the convention that  
  $n $ in eq.(\ref{quacon}) should be counted from 1 and  $\eta_{np}$  should 
be confined to
  the interval between $+\pi/4$ and $-3\pi/4$.\footnote{Note that with $n=1$ and  $\eta_{np} = 0$  the eq.(\ref{phicon})  is well satisfied, however it is not satisfied with $n=0$ and  $\eta_{np} = 0$,  
   since  $s_\omega(t_0)$ is always positive. The periodicity of eq.(\ref{phicon}) assures that the same eigenfunction is obtained with $n=1$ and $\eta_{np}=0$  as with   $n=0$ and $\eta_{np}=\pi $.}
The values of $\eta_n$ and the corresponding eigenfunctions, used later in the fit, are not limited to this interval. They are obtained from the periodicity of  $\eta_n$, i.e. by adding (or subtracting) multiples of $\pi$ on both sides of eq.(\ref{phicon}).   
   In the following we will label the eigenvalues and eigenfunctions with $n \ge 1$ and denote the $n$ dependent  phase   $\eta_{np}(n)$ simply by   $\eta_{n}$.

\begin{figure}[htbp]
\centerline{
\includegraphics[width=8.9cm,angle=0]{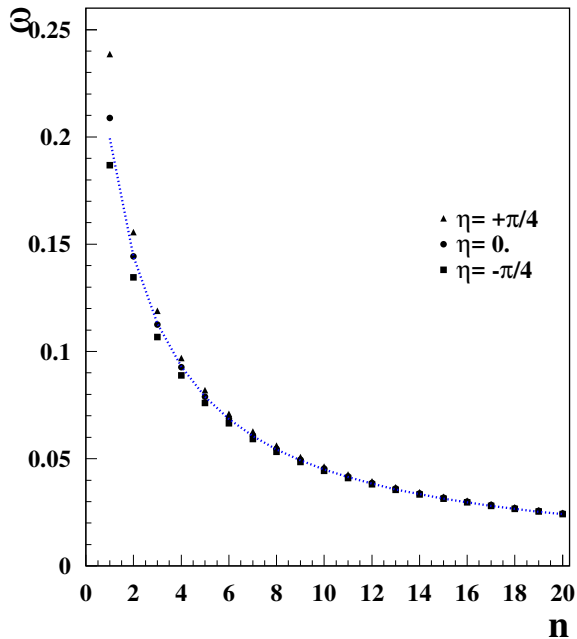}
}
\caption{ Eigenvalues $\omega_n$ determined in NLO for three fixed non-perturbative phases, $\eta_{n}$. The dotted line shows a simple parametrisation described in the text.    }
\label{omd} \end{figure}

\begin{figure}[htbp]
\centerline{
\includegraphics[width=8.9cm,angle=0]{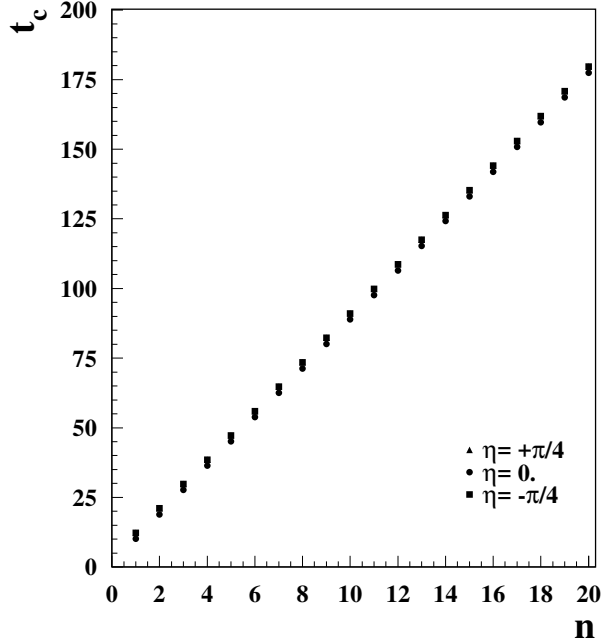}
}
\caption{ The critical momenta $t_c$  determined in NLO for three fixed non-perturbative phases, $\eta_{n}$. $t_c =\ln k_c^2/\Lambda_{QCD}^2$ with $ \Lambda_{QCD} = 275$ MeV.}
\label{tc} \end{figure}

In Fig.~\ref{omd} we display the eigenvalues $ \omega_n$ obtained from eq.(\ref{pert13}), using three different non-perturbative  phases, $\eta_{n} = 0, \pi/4, -\pi/4$.
The dotted line shows,  as an example the $\eta_{n} = 0$ case,  that the dependence of $ \omega_n$ values from $n$ (for $n > 1$) can be simply parametrised by 
\beq \omega \ = \ \frac{A}{n+B} \label{ompar}\ , \eeq
as noticed already in~\cite{lipatov86}.
For $\eta_{n} = 0$ we found in NLO, that A =   0.52223, B=   1.62001. Since we apply this parametrisation below to describe data we recall  its derivation 
given in  ref.~\cite{lipatov86}.
In LO we can integrate $s_\omega(t_0)$  by parts
\beq s _\omega(t_0) \ = \ \int_{t_0}^{t_c} \nu_\omega(t') dt'  =  - \nu_\omega(t_0) t_0 + \frac{1}{\betabar \omega} \int_0^ {\nu_\omega(t_0)}  \chi_0(\nu')d\nu'  \label{sompart}, \eeq
where  in the last step we used the LO relation $t =  \chi_0(\nu)/  \betabar  \omega $. For $\omega$ values approaching 0, we have 
 \beq \chi_0\left(\nu_\omega(t)\right) \ =  \ \frac{\omega}{\alphabar(t)} \rightarrow  0
 \label{freq} \eeq 
Therefore, for small $\omega$ and small $t_0$,  $\nu_\omega	$ is  quickly approaching its asymptotic value,  $\nu_0$, with  $\chi_0(\nu_0) = 0$.
In this limit $\int_0^ {\nu_\omega(t_0)}  \chi_0(\nu')d\nu'  $ and $\nu_\omega(t_0)$  become independent of $\omega$ and eq.(\ref{phicon}) implies that  
 \beq n\pi \ = a +  \frac{b}{\betabar \omega}   + \frac{\pi}{4} -  \eta_{n} , \label{par1} \eeq 
where $ a,b$ are constants independent  of  $\omega$. This leads to the relation~\ref{ompar}.  In NLO this relation is  satisfied already for $n \ge 2$,  since  
$\nu_\omega(t_0)$ is less dependent on $\omega$ than in LO. The relation~\ref{ompar} indicates also that  for large $n$,  $t_c =  \chi_0(0)/  \betabar  \omega_n $ should grow almost linearly  with $n$.  This is also a feature of the NLO computation, see Fig.\ref{tc}. The value of $t_c$ is related to the value of the critical momenta $k_c$  by $t_c =\ln k_c^2/\Lambda_{QCD}^2$ with $ \Lambda_{QCD} = 275$ MeV. 

\begin{figure}[htbp]
\centerline{
\includegraphics[width=17.0cm,angle=0]{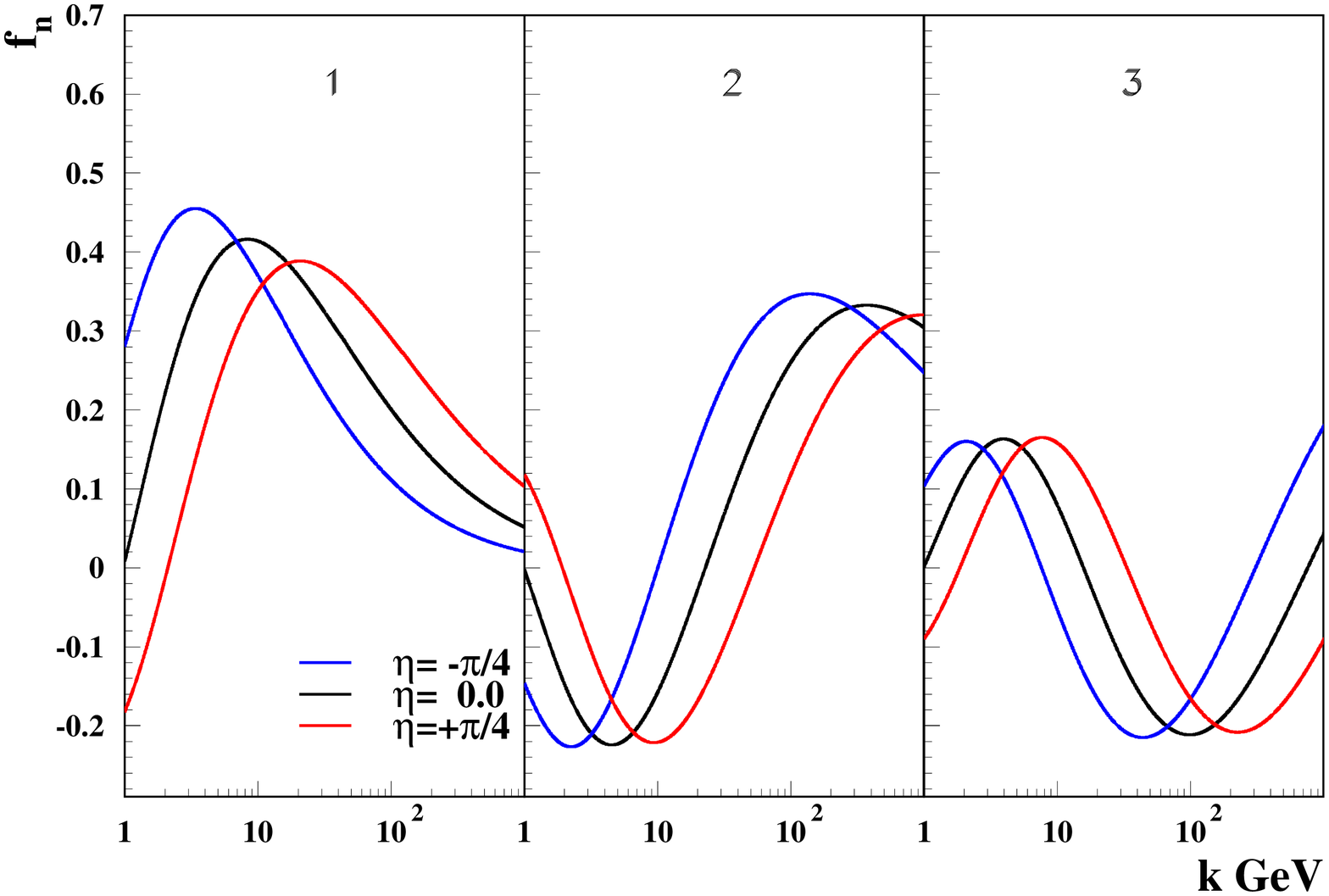}
}
\caption{ The first three eigenfunctions  computed for three fixed non-perturbative phases, $\eta_{n}$.    }
\label{3efs} \end{figure}

 In Fig.~\ref{3efs} we show as example  the first three different eigenfunctions 1,2 and 3, computed from eqs.(\ref{normeigen}) and (\ref{pert13}),   at  phases   $\eta_{n} = 0, \pi/4, -\pi/4$. 

\section{Application to data}


To apply the BFKL Green Function to data,  we express  the low-$x$ structure function of the proton, $F_2(x,Q^2)$, in terms of the discrete BFKL eigenfunctions by 
\begin{equation}
F_2(x,Q^2) \ = \ \int_x^1 d\zeta \int \frac{dk}{k} \Phi_{\gamma}(\zeta,Q,k)
  x g\left(\frac{x}{\zeta},k\right),
 \label{f2-expr}
\end{equation}
where $x g\left(\frac{x}{\zeta},k\right)$ denotes the unintegrated gluon density
\begin{equation}
x g(x,k) \ = \  \int \frac{dk^\prime}{k^\prime} \Phi_p(k^\prime)
 \left(\frac{k^\prime \, x}{k}\right)^{-\omega_n} k^2   \left( \sum_n  f^*_{\omega_n}(k^\prime) f_{\omega_n}(k)  
 +  \int_{-\infty}^0 d\omega x^{-\omega}  f_{-|\omega|}(t) f_{-|\omega|}(t^\prime)
 \right)
\label{un-glu}
\end{equation}
and $\Phi_p(k)$ denotes the impact factor that describes how proton couples to the  BFKL amplitudes at zero momentum transfer. The impact factor, $\Phi_{\gamma}(\zeta,Q,k)$,  which describes the coupling of the virtual photon to the eigenfunctions is given in \cite{KMS}; the dependence on $\zeta$ reflects the fact that beyond the leading logarithm approximation, the longitudinal momentum fraction, $x$, of the gluon differs from the Bjorken-value, determined by $Q^2$.   $\Phi_{\gamma}(\zeta,Q,k)$ of Ref.~\cite{KMS} is determined taking into account kinematical constraints allowing for non-zero quark masses.
The $(k'/k)^{\omega_n}$ factor arises
 from a mismatch between the ``rapidity'', $Y$, of the forward gluon-gluon
scattering amplitude used in the BFKL approach
 $$ Y \ = \ \ln\left(\frac{s}{k k^\prime}\right) $$
and the logarithm of Bjorken $x$, which is given by
 $$ \ln\left(\frac{1}{x}\right) \ = \  \ln\left(\frac{s}{k^2}\right). $$
This ambiguity has no effect in LO but in NLO it can be compensated by
 replacing the LO characteristic function $\chi_0(\nu)$
 by $\chi_0(\omega/2,\nu)$, which modifies  the NLO characteristic function $\chi_1$ (see Appendix A). 

 The proton impact factor is determined by the confining forces. It is therefore barely known, besides the fact that it should be concentrated at the values of $k < {\cal O}(1)$ GeV.  We  use here a simple parametrisation in the form
\beq \Phi_p(\mathbf{k}) \ = \ A \, k^2 e^{-b k^2} \label{impact}, \eeq
which vanishes as $k^2 \, \to \, 0$, as a consequence of colour transparency and is everywhere positive. The value of $b$ should be around 13 GeV$^{-2}$, i.e of the inverse square of $\Lambda_{QCD} = 275$ MeV. This is much higher than the value of $b$ determined from data for the proton form factor, $b \approx 4 $ GeV$^{-2}$.   Since the range  of the proton impact factor is much smaller than the oscillation period of the BFKL eigenfunctions we do not expect that the results should have substantial sensitivity to a value of $b$.  Therefore we performed the investigation assuming two very different values of the impact factor, $b= 10$ and $b=20$ GeV$^{-2}$, corresponding to $\Lambda_{QCD} \approx 320$ or 220 MeV. We also used, as a check, an extreme proton impact factor, $  \Phi_p(\mathbf{k})  = A  \, \delta(k-k_0)  $.


\subsection{Properties of HERA data}
The HERA $F_2$ data in the low $x$ region can be simply parametrised by $F_2 = c\  (1/x)^\lambda$, with the constants $c$ and $\lambda$ being  functions of $Q^2$, see e.g.~\cite{Cald}.  As $Q^2$ increases from 4 GeV$^2$ to 100 GeV$^2$   $\lambda$ changes from about 0.15 to 0.3.  The BFKL evaluation of $F_2$, which assumes that $\eta_n$ is independent of $n$, would predict that $\lambda$ is a constant, $independent$ of $Q^2$ with   $\lambda \approx \omega_1$,
since it is the  first pole which  dominates $F_2$, when the  value of $\eta_n$ is fixed. Therefore, the only way that $\lambda$  can depend on  $Q^2$ is if
  the infrared phases,  $\eta_n$, depend on  $n$. Otherwise,  the predicted  value of $\lambda$ will be  about 0.25,  $independent $ of $Q^2$   (see Fig.\ref{omd}) in clear contradiction with HERA data.

 The fits utilize the highest precision HERA data \cite{hiH1ZEUS}   given in terms of reduced cross sections from which we extracted  the $F_2$ values, using  the  assumption that $F_L$ is proportional to $F_2$. We also limit
   the $y$ range in order  to avoid possible complications of a larger
   contribution from  $F_L$ (see e.g.~\cite{Cald}). 
 Since we are focusing  on the comparison with the $F_2$ measurements,
  we only use  the  920 GeV data set of~\cite{hiH1ZEUS}. We also limited the comparison with data to the  region   $x < 0.001$ and $Q^2 > 6$ GeV$^2$ since the BFKL equation is valid at very low $x$ only. The $Q^2$ cut was chosen to be relatively high to avoid any complications due to possible saturation corrections~\cite{LuszKow}. The number of experimental points used for fits  was then $N_p=51$.  (It represents around 1/3 of the whole low $x$ data sample, defined as $x< 0.01, \ \ Q^2 > 3$ GeV$^2$).
 
 For this investigation  we have taken the uncorrelated errors, obtained by adding in quadrature all the correlated errors of ref.~\cite{hiH1ZEUS}. From the data analysis of ref.~\cite{LuszKow} we know that  the uncorrelated errors overestimate the error sizeably, so that the     $\chi^2/N_{df}$ of a good fit should be around  0.7, instead of about 1 as in  case of correlated errors (see also~\cite{HERAFitter}).   
 
  \subsection{Boundary condition}
 The major challenge in confronting the BFKL predictions with data is the
 determination of the infrared boundary condition. i.e. finding the relation between the infrared phases, $\eta_n$ and the eigenfunction number  $n$ which generates a precise description of the data.
At the beginning we tried to parametrise $\eta$ as a function of $n$, using polynomial or other functional dependences. This failed because we were not able to find any functional dependence which would lead to $\chi^2 < O(500)$. In the next step we tried to find a set of  $\eta_n $ (with $n=1,2,3...10$) values using only some assumptions of local continuity. This was essentially a 10 parameter fit, with some limitations. After a longer search, using permutational methods to avoid any pre-conceptional bias on the form of  $\eta - n$ relation,  we found  a set of 10 $\eta_n$  values which gave  an acceptable  $\chi^2 \approx 40$. Studying this set we noticed that it can be well parametrised by an  $\omega - n$ relation, similar to 
eq.(\ref{ompar}),
\beq \omega \ = \ \frac{A}{n+B} + C \label{omparc}\ , \eeq
with a value of  $C$ which is very small, but nevertheless non-zero.
 The $\eta_n $ values were then obtained from eqs.(\ref{phicon}) and (\ref{quacon}),  by  
 \beq 
  \eta_{n}  \ = \ s_{\omega_n}(t_0) + \frac{\pi}{4} -n\pi  .  \label{etan} \eeq 
The parameters $A$,  $B$ and $C$, together with  $\eta_{neg}$, the phase of the negative omega contribution,   were considered as  free parameters of the fit, which  we call in the following the ABC-Fit. In addition to these four parameters  the overall normalisation  was also fitted to  data. 

\begin{figure}[htbp]
\centerline{
\includegraphics[width=16.0cm,angle=0]{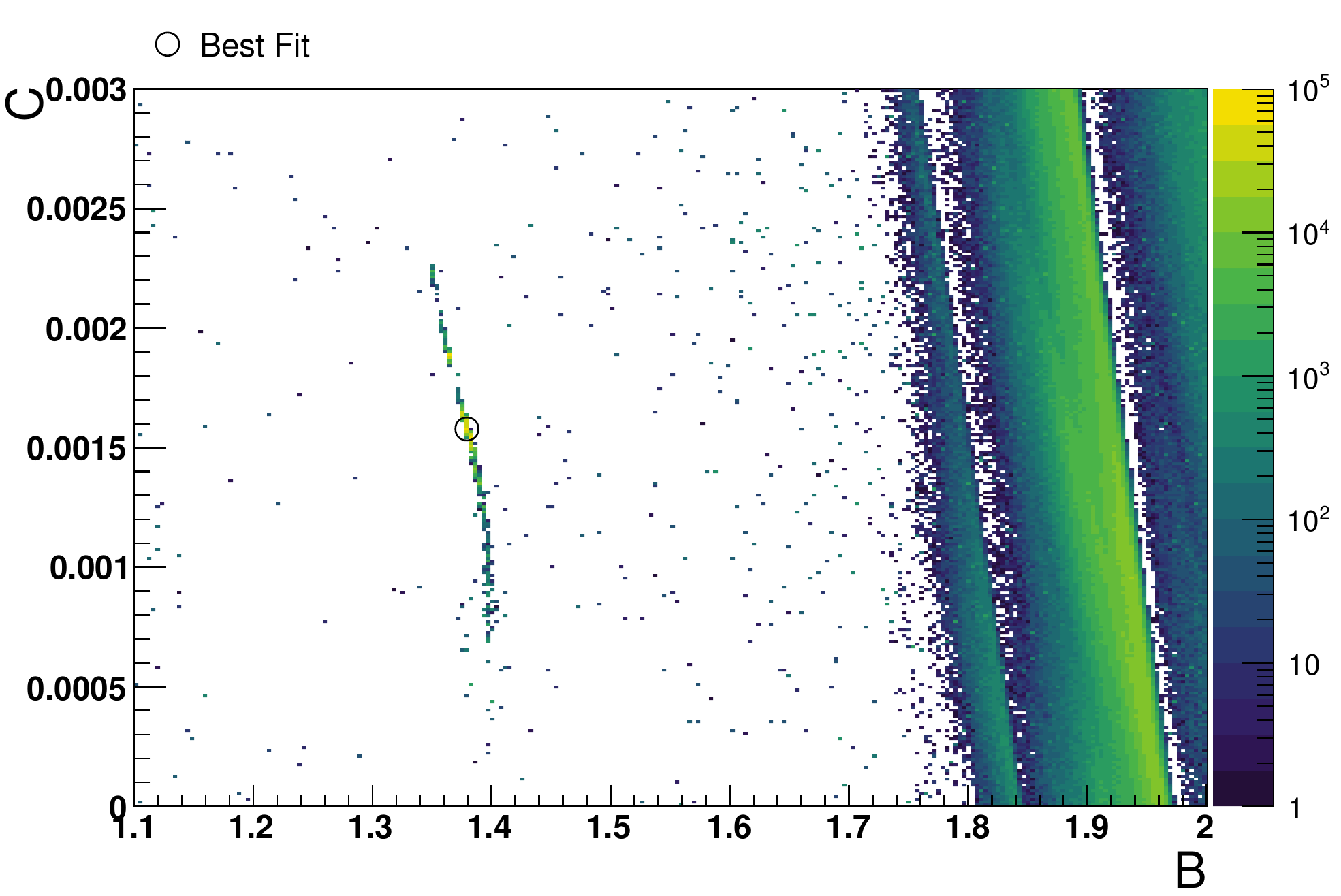}
}
\caption{ Probability density of the $ABC$ fit as a function of the B and C parameters. The legend shows the probability  scale in arbitrary units   }
\label{ABC} \end{figure}

\begin{figure}[htbp]
\centerline{
\includegraphics[width=16.0cm,angle=0]{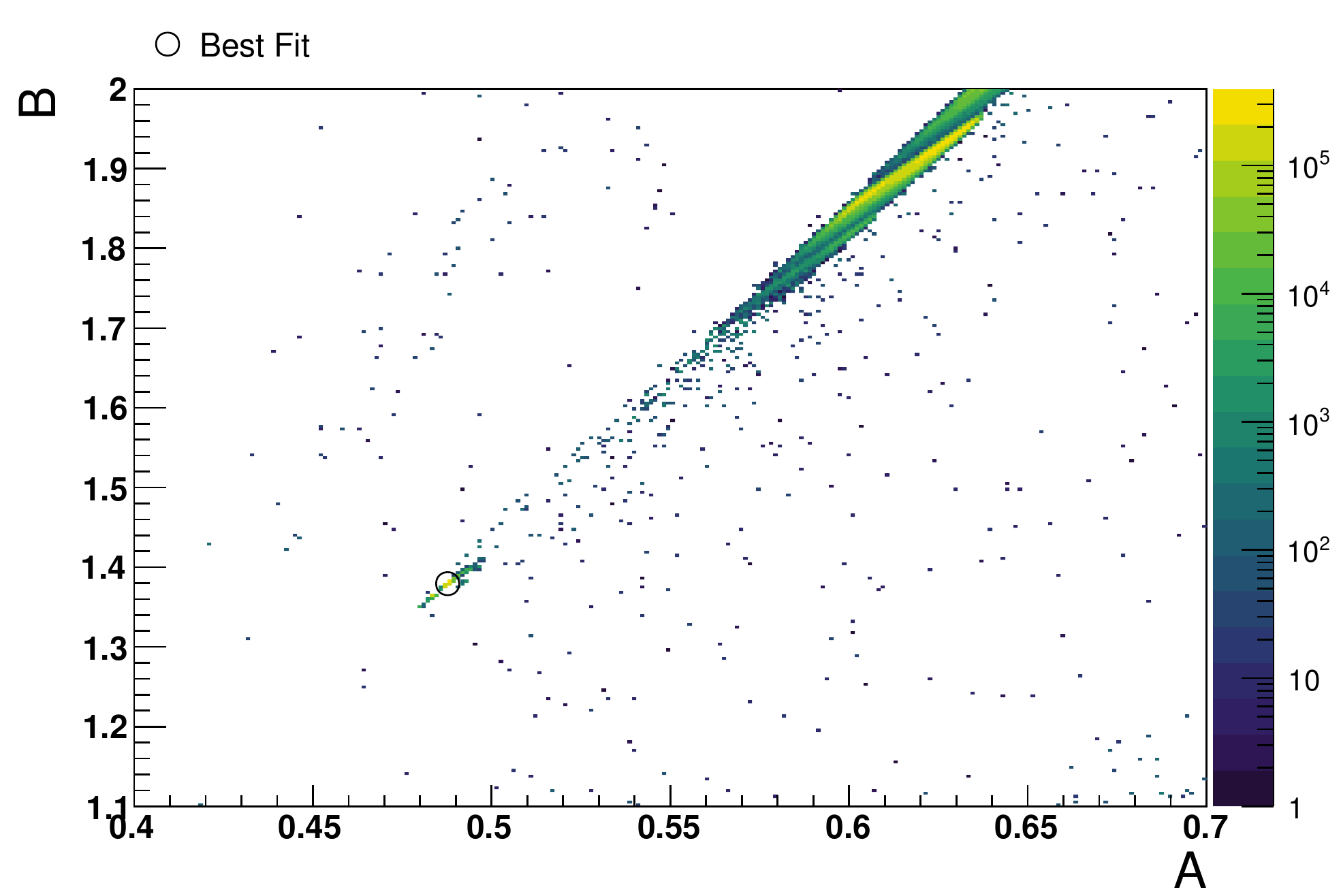}
}
\caption{ Probability density of the $ABC$-Fit as a function of the B and A parameters. The legend shows the probability  scale in arbitrary units   }
\label{ABC-1} \end{figure}

As we observed that the system was exhibiting a multitude of local
optima, we used the Bayesian Analysis Toolkit (BAT)~\cite{BAT}  to find
the global optimum. BAT generates samples in  parameter space
via Markov chain Monte Carlo (MCMC), distributed according to
the posterior probability of the parameters. The best fit value
is the parameter set with the highest posterior probability, corresponding to
the lowest $\chi^2$-value.
 Fig.~\ref{ABC} shows a marginalised  distribution of the $ABC$-Fit, for the variables, $B$ and $C$. 
The regions  of higher probability are shown as coloured areas, with probability increasing as the colour changes from blue over green to yellow. 
The small circle shows the position of the best fit, given in Table 1. The complicated structure of the probability distribution is also seen as a function of $A$ and $B$ variables, (see Fig.~\ref{ABC-1})  

 Fig.s~\ref{ABC} and~\ref{ABC-1}  show that the distribution of  probability has a  complicated structure;  there are several extended regions of  higher probability, which  are completely disconnected.   In this situation the usual fitting methods, based on MINUIT, work poorly, since they   assume a 
 smooth increase in probability towards the real minimum.
 
   Using the BAT together with the above parametrisation we found an excellent agreement with data,  $\chi^2/N_{df} \approx 33/46$.    
We performed this fit for several specific values of the parameters $b$ of $\Phi_p$ and found that the $\chi^2$ values were the same, within the computational precision of the fit, $\Delta \chi^2 = \pm 1$. For each value of $b$ the values of the fit parameters, $A,B,C$ and $\eta_{neg}$,   were somewhat different and compensated  the change of $b$ (see for example Table 1).  
\begin{table}[h]
\begin{center}
 \begin{tabular}{||c|c|c|c|c|c|c|c|||} \hline 
  $b$ (GeV$^{-2}$)  & 10 & 20   \\ \hline
   $A$        &  0.48771  &  0.47905    \\  \hline
   $B$ &   1.37933   &  1.34020    \\  \hline
   $C$ &   0.001578   &  0.002424   \\  \hline
    $\eta_{neg}$ &  -0.0754  &   -0.0518 \\ \hline
    $\chi^2$ &  32.9  &  33.1   \\ \hline 
 \hline \end{tabular}
\caption{ Results of the ABC-Fit to 51 data points with $x<0.001$ and $Q^2> 6$ GeV$^2$. }
\end{center}   \end{table}
The values of the $A$ and $B$ parameters are  in the usual range, $A \approx 0.5$, $B \approx 1.5$, similar to the values at fixed phase,  $\eta$, (see eq.(\ref{ompar}) and below).   The third parameter, $C$, is very small, O($10^{-3}$), i.e. much smaller than the value of the smallest eigenvalue, $\omega_{20} \approx 0.025$, used in the fit. 
 
  In spite of the fact that $C$ is very small, it is impossible to put its value to  zero   without seriously deteriorating the  quality of the $ABC$ fit (to  $\chi^2 \approx 150$).  In   standard QCD we should expect  $C$  to be zero
 so that $\omega_n \rightarrow 0$ when $n \rightarrow \infty$, as in the LO calculation discussed above.
However,  we noticed, that the parameter $C$ can to be set to zero if we let  $\eta_1$, the phase  of the first eigenfunction, to be  a free parameter, instead of $C$.  The fits obtained in this way are of the same quality as the $ABC$ fits, they have  however  an unexpected property; the value of the $\eta_1$ parameter is always chosen such that  the first eigenfunction decouples (or nearly decouples) from the proton. This means that its overlap with the proton form-factor becomes zero (or nearly zero), independent of the choice of $b$. 
Therefore, we determined the phase $\eta_1$  solely from the requirement that the first eigenfunction should be orthogonal to the proton impact factor 
(in this way  the parameters A and B are correlated,   for a given impact factor, with the value of the phase $\eta_1$).
We call this fit  the AB-Fit and give its   results in Table 2, for two values of $b$ as example.\footnote{ The values of $\eta_1$ at the decoupling point, in the AB fit, are
 $\eta_1 =  0.0707$  for the $b= 10$ and  $\eta_1 =0.0503 $ for the $b= 20$ GeV$^2$ case.}
In the  AB-Fit the first eigenfunction is not used since it is decoupling from the proton.
In addition, we note  that an  approximate decoupling happens also in the ABC-Fit, where the contribution of the first pole is much smaller than that of the second one, by more than a factor of 10.  Finally we note that in fits of Table 1 and 2  we used 20 eigenfunctions,  to see the convergence (see below).   


\begin{table}[h]
\begin{center}
 \begin{tabular}{||c|c|c|c|c|c|c|c|||} \hline 
  $b$ (GeV$^{-2}$)  & 10 & 20   \\ \hline
   $A$        &  0.51844  &  0.51913    \\  \hline
   $B$ &   1.58697   &  1.58657    \\  \hline
   $\eta_{neg}$ &  -0.0911  &   -0.0550  \\ \hline
    $\chi^2$ &  33.9  &  33.3   \\ \hline 
 \hline \end{tabular}
\caption{ Results of the AB-Fit to 51 data points with $x<0.001$ and $Q^2> 6$ GeV$^2$. }
\end{center}   \end{table}

\begin{figure}[htbp]
\centerline{
\includegraphics[width=16.0cm,angle=0]{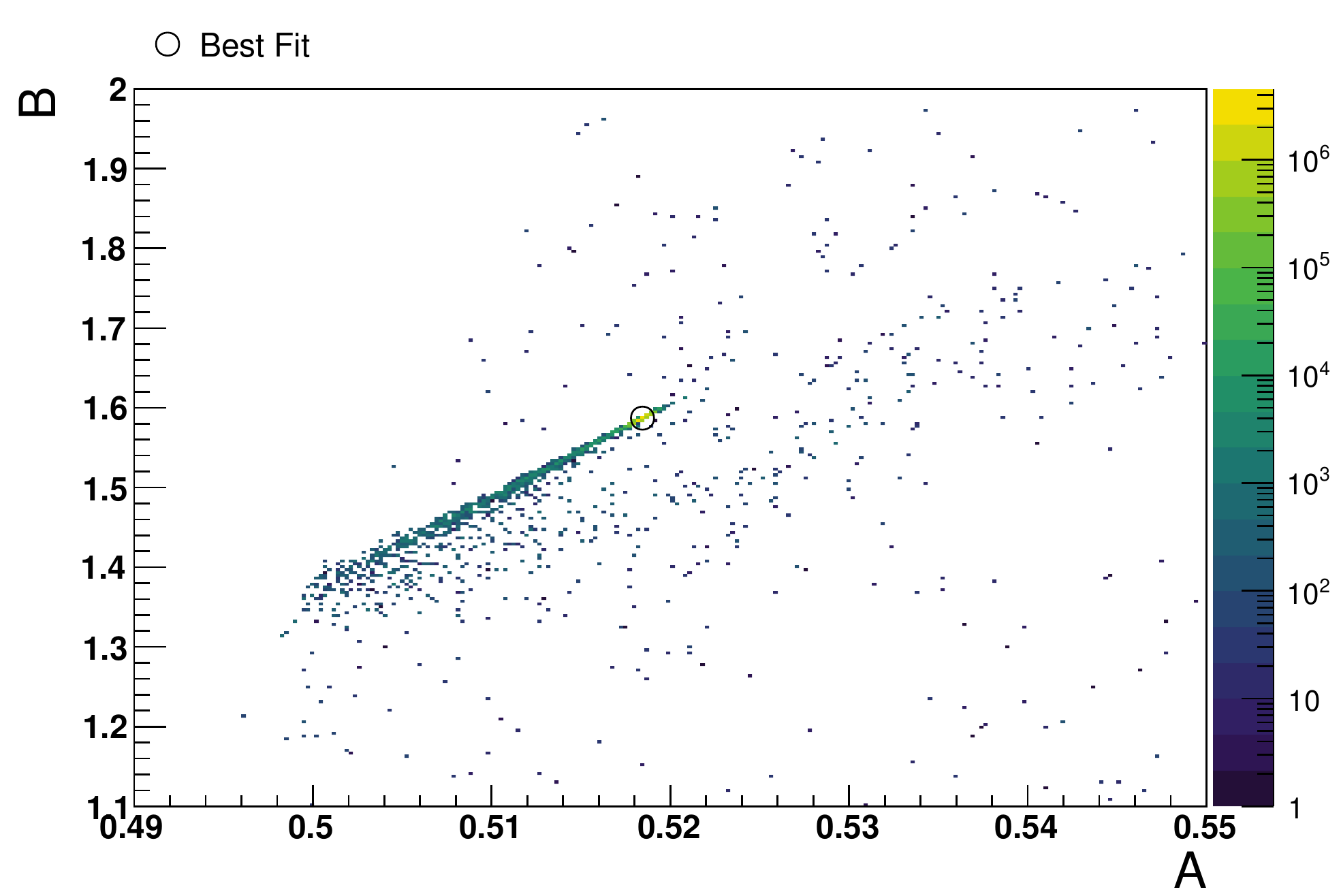}
}
\caption{ Probability density of the $AB$-Fit as a function of the B and A parameters. The legend shows the probability  scale in an arbitrary units    }
\label{AB} \end{figure}

The assumption of the decoupling of the first eigenfunction, together with the  AB-relation 
of eq.(\ref{ompar}), leads to a much simpler probability  structure, (see Fig.~\ref{AB}), with a steady increase of probability towards one minimum, i.e.,  without a multitude of local minima.

\begin{figure}[htbp]
\centerline{
\includegraphics[width=12.0cm,angle=0]{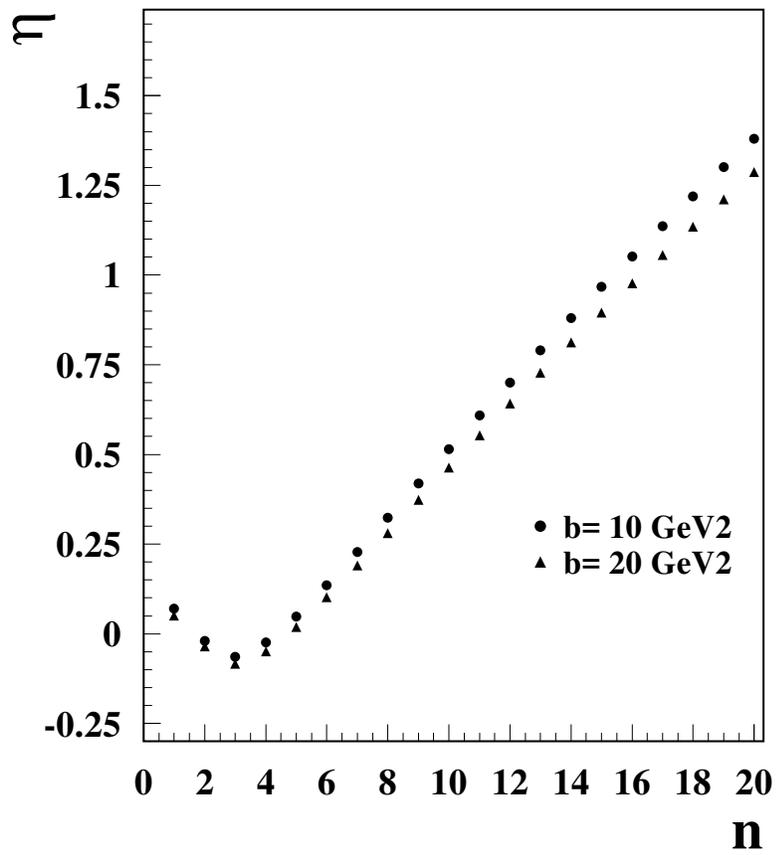}
}
\caption{   $\eta - n$ relation as computed from the parameters $A,B$ of the AB-Fit.  }
\label{etaom} \end{figure}

\begin{figure}[htbp]
\centerline{
\includegraphics[width=12.0cm,angle=0]{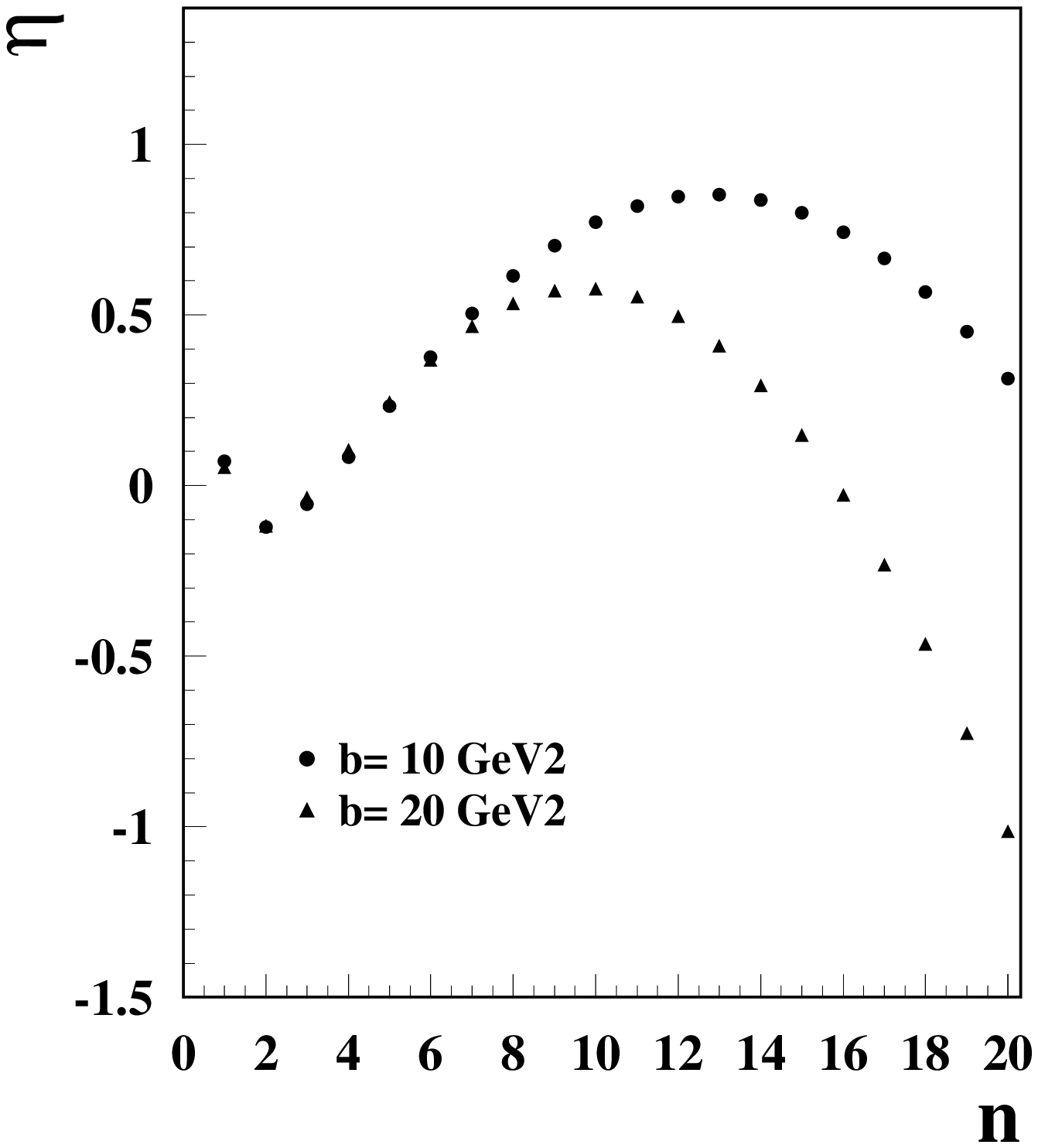}
}
\caption{   $\eta - n$ relation as computed from the parameters $A,B,C$ of the ABC-Fit.  }
\label{etaabc} \end{figure}

 In Fig.~\ref{etaom} we show the $\eta - n$ relation as computed from the parameters $A,B$ of the AB-Fit for two values of $b$. Note that  $\eta - n$ relation is visibly different in the two cases,  although the parameters $A,B$ differ by a fraction of per mill only.  In Fig.~\ref{etaabc} we show the same relation 
 as computed from the parameters $A,B,C$ of the ABC-Fit for the same two values of $b$. Note that the 
  $\eta - n$ relation is simpler in the AB-Fit than in the ABC-Fit. 

In general, we observe that the AB and ABC parameterisations are characterised by a high sensitivity to the 
 values of $\omega$. The values of the parameters $A,B$ for the case of constant $\eta$, given below eq.(\ref{ompar}), differ only by  about a percent from the values in Table 2,  and yet produce a very different   $\eta - n$  relation. 
 A fit to data with constant $\eta$ would give  $\chi^2 \approx 3000$!

\subsection{Fit results}

\begin{figure}[htbp]
\centerline{
\includegraphics[width=12.0cm,angle=0]{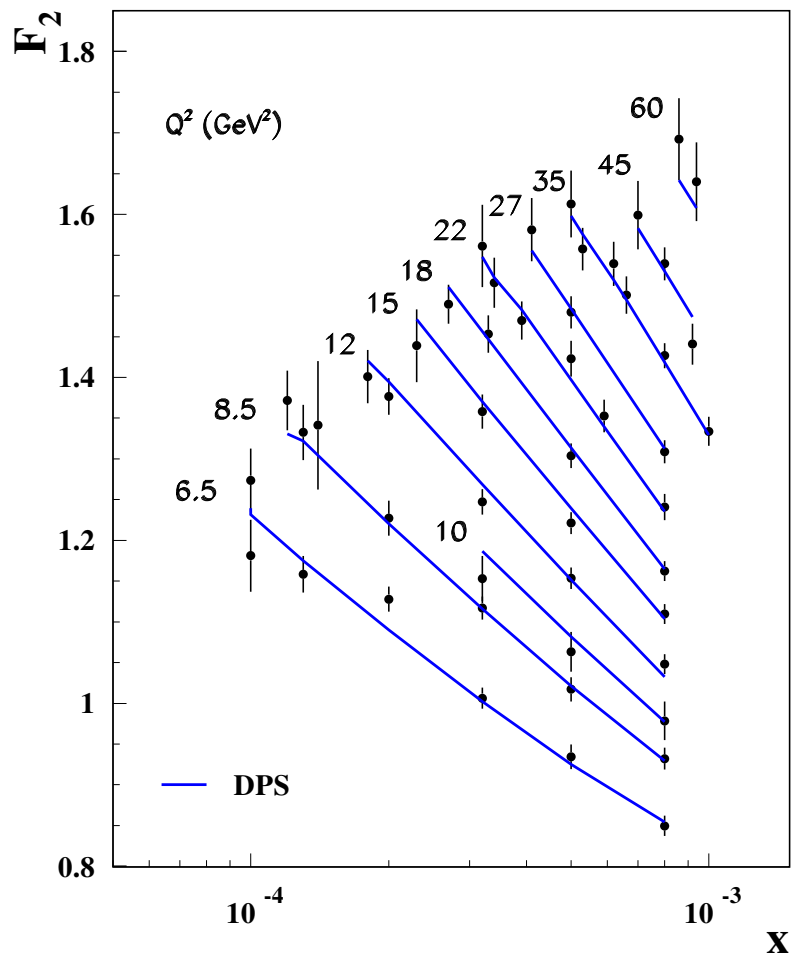}
}
\caption{ Comparison of the $AB$ fit results with data   }
\label{globfit} \end{figure}

\begin{figure}[htbp]
\centerline{
\includegraphics[width=11.0cm,angle=0]{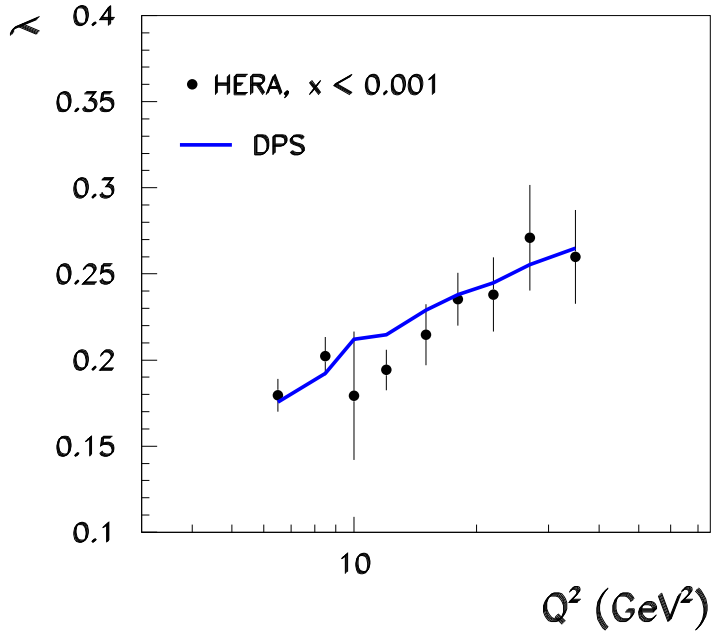}
}
\caption{ Comparison of the $\lambda$ parameter, obtained in the $AB$ fit,  with data.   }
\label{lam} \end{figure}

\begin{figure}[htbp]
\centerline{
\includegraphics[width=9.0cm,angle=0]{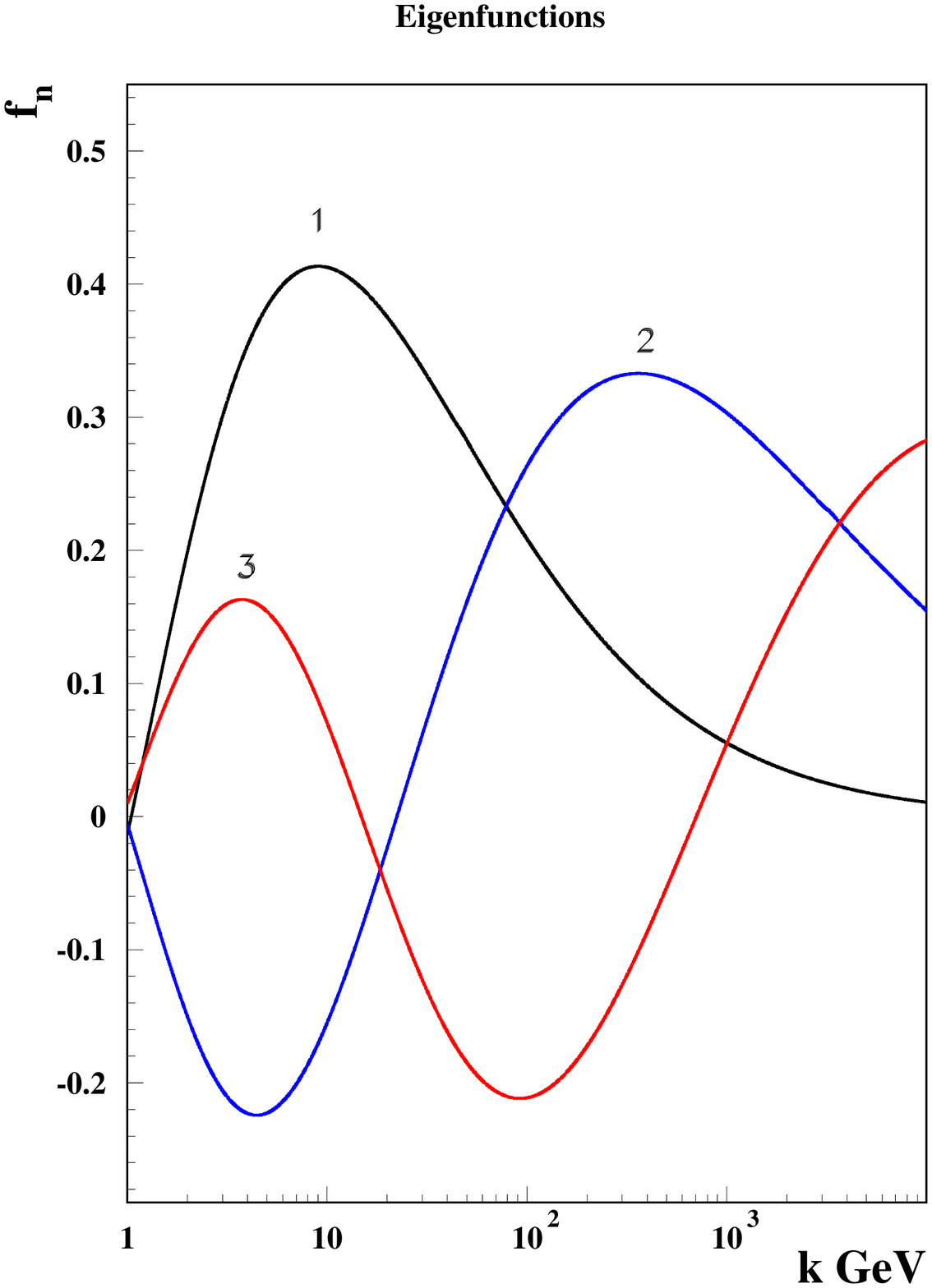}
\includegraphics[width=9.0cm,angle=0]{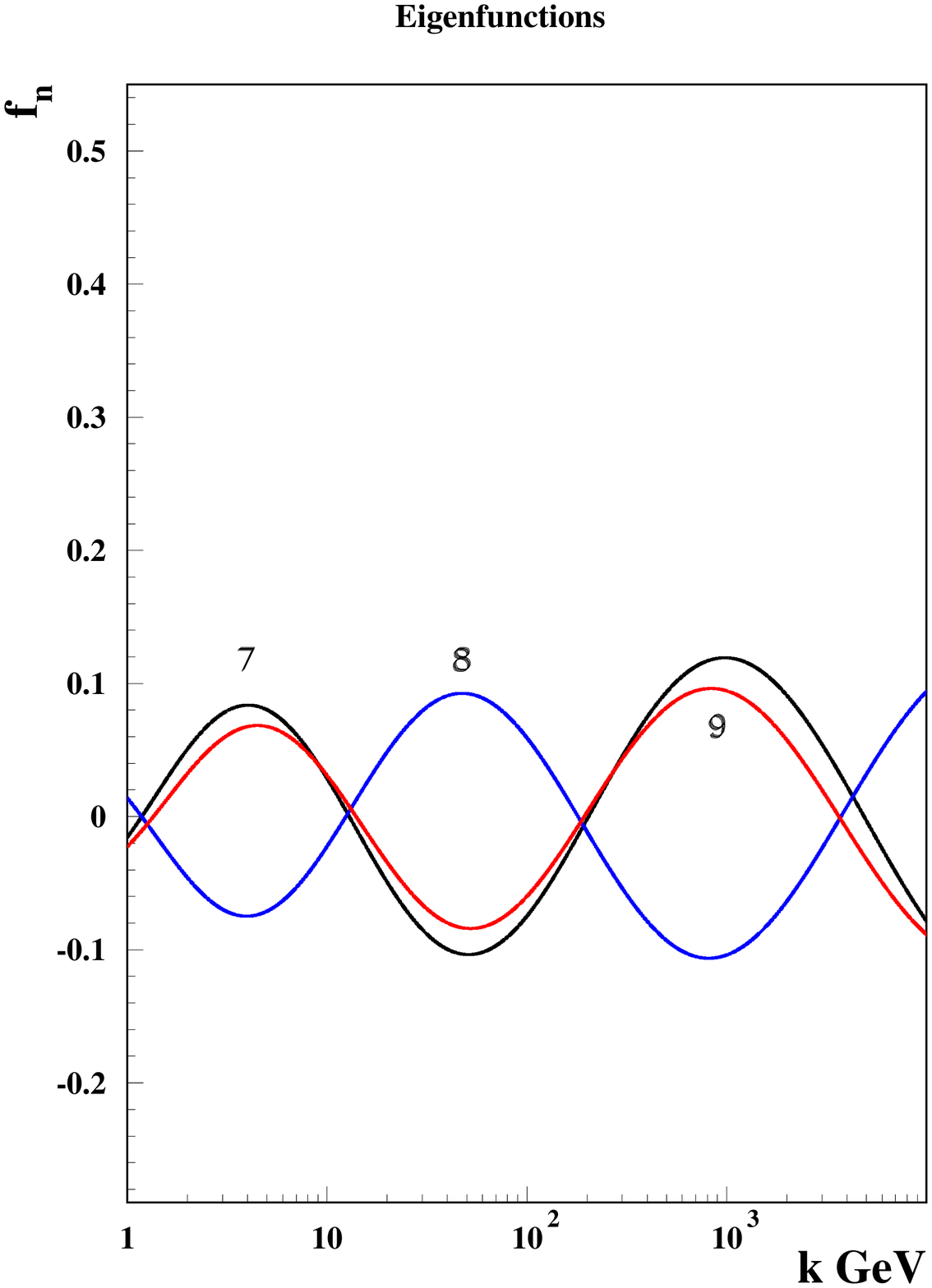}
}
\caption{   Eigenfunctions 1, 2, 3, and 7, 8 and 9  in the  $k$ region accessible to experiments. The eigenfunctions are plotted with the $\eta_n$ phases given by the $AB$ fit, performed with the $b$ value of the proton impact factor equal to $10$ GeV$^{-2}$ . The first eigenfunction is plotted with the phase  which decouples it from the proton. }
\label{eigen123} \end{figure}

\begin{figure}[htbp]
\centerline{
\includegraphics[width=12.0cm,angle=0]{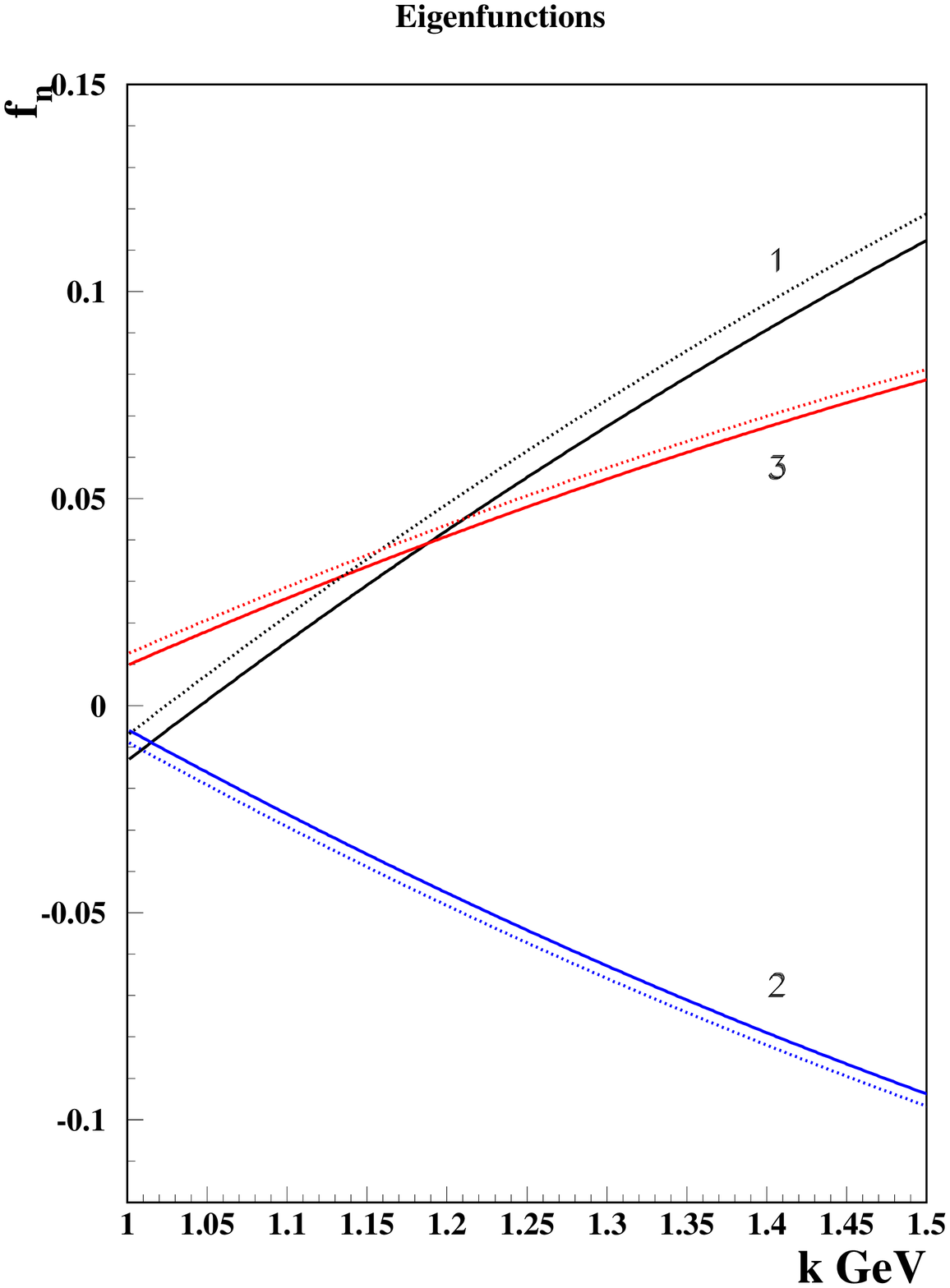}
}
\caption{   Eigenfunctions 1, 2, 3,   in the  $k$ region close to $k_0$. The eigenfunctions are plotted with the $\eta_n$ phases given by the $AB$ fit. The first eigenfunction is plotted with the phase which gives zero overlap with the  proton impact factor. The fits were performed with the $b$-value of  the proton impact factor given by $10$ GeV$^2$ (full lines) and $20$ GeV$^2$ (dotted lines)}
\label{eigenblow} \end{figure}

In Fig.~\ref{globfit} we show the comparison of the AB-Fit results with data (with $b=10$ GeV$^2$). Fig.~\ref{globfit} shows a very good agreement,  corresponding to the excellent $\chi^2$ value. The results obtained with different choices of parameter $b$, or with ABC-Fit, would look the same in this figure.

The  BFKL Green Function, determined in our approach, is able to describe the $Q^2$  dependence of the data, given by the $F_2$ values or by the  slope $\lambda$,  although neither  the  eigenvalues nor the AB(C)-parameters are $Q^2$ dependent.
In Fig.~\ref{lam} we show  the comparison of the $\lambda$ parameter obtained from the AB-Fit  with data. The  $\lambda$ parameter  
was determined in the very low $x<0.001$ region and in the $Q^2$ range between 6.5 and 35 GeV$^2$.
 The $Q^2$ dependence enters indirectly because  the eigenfunctions  depend on the transverse momentum $k$, which in the convolution with the photon impact factor, leads to a $Q^2$ dependence .

 
 \subsection{Discussion of the phase tuning mechanism}
 
 The choice of  the $\omega - n$ relation determines the set of phases $\eta_n$ which tune the contributions of the individual eigenfunctions to describe the data.
 To see how this happens we display in Fig.~\ref{eigen123}  the  eigenfunctions 1, 2, 3, and  as an example of subleading ones the eigenfunctions   7, 8, 9, as a  function of  $k$. The eigenfunctions are plotted with the $\eta_n$ phases, for $n \ge 2$, given by the AB-Fit. The first eigenfunction has the phase $\eta_1$, which  suppresses its overlap with the proton impact factor.
  The figure  shows that the leading eigenfunctions  2 and 3 have the values $f_n (k_0) \approx 0$, whereas the eigenfunctions   7, 8 and 9, have the values at  $k_0$ which are substantially different from zero. 
  
To see more precisely how the phases determine the overlaps, we display in Fig.~\ref{eigenblow}  the eigenfunctions 1, 2 and 3
in the region close to $k_0$, for the fits with $b=10$ (full lines) and~20 (dotted lines) GeV$^2$.  We see that, in both cases, the eigenfunction 1 starts negative at 
$k_0 = 1$~GeV but then crosses zero at  $k_0 \approx 1.05$ and becomes positive. This small negative region is sufficient to
 suppress the overlap with the proton impact factor and effectively cancel its contribution to $F_2$.  
The eigenfunction 2 and 3 do not cross zero, and in both cases the overlap with the proton and DIS impact factors have the same signs. They give, therefore, large contributions to $F_2$. The contributions of the subleading eigenfunctions 7, 8 and 9 are also significant because $\eta_n$ values are substantially different than zero,  $\eta_7 = 0.23$, $\eta_8 = 0.32$ and $\eta_9 = 0.42$. This leads to  large overlaps with the proton and photon impact factor, but in this case they have have opposite signs.  
Their contributions to $F_2$  are therefore relatively large and have negative sign  so  that they
  can generate a $Q^2$ dependence in  the slope $\lambda$.

\begin{figure}[htbp]
\centerline{
\includegraphics[width=10.0cm,angle=0]{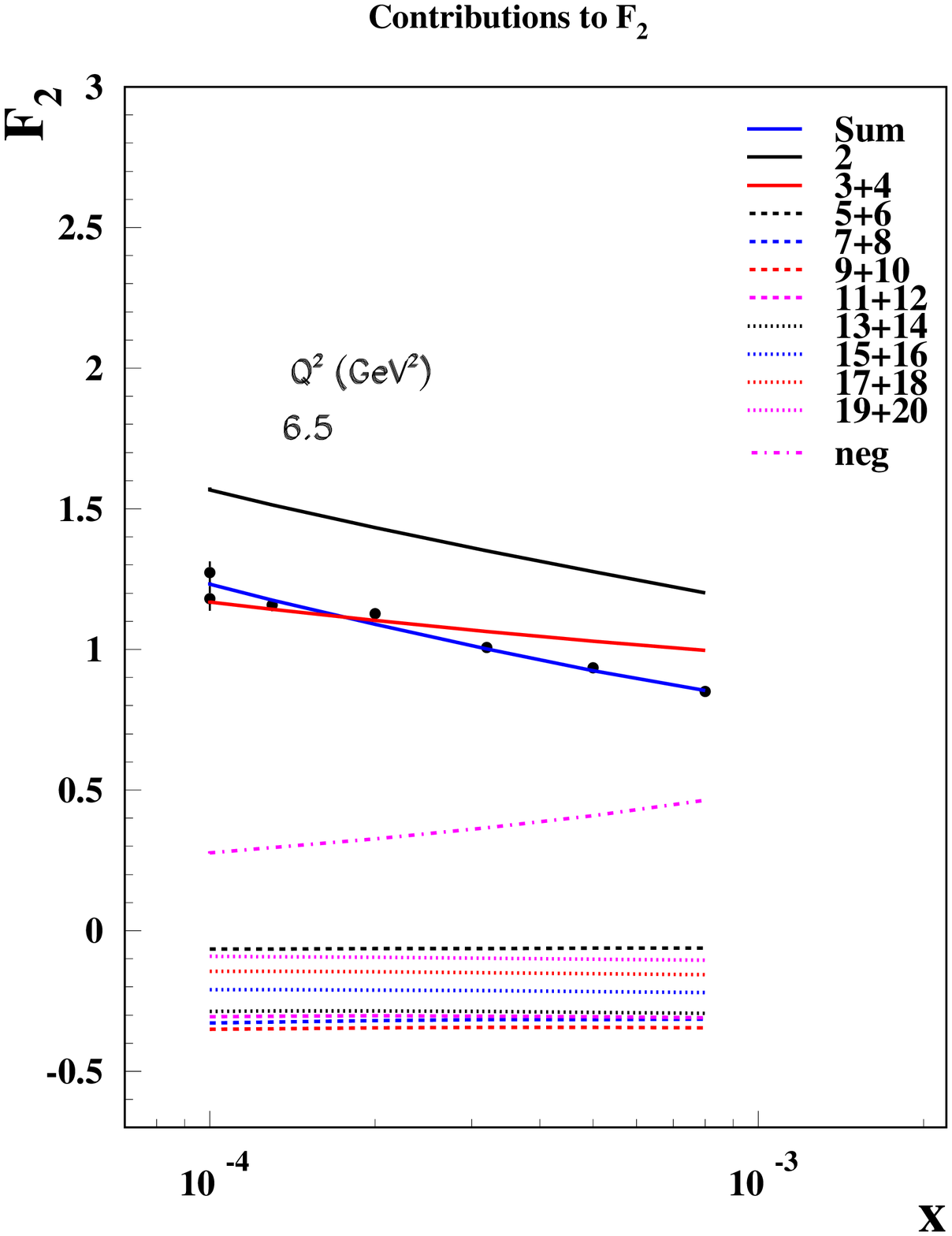}
\includegraphics[width=10.0cm,angle=0]{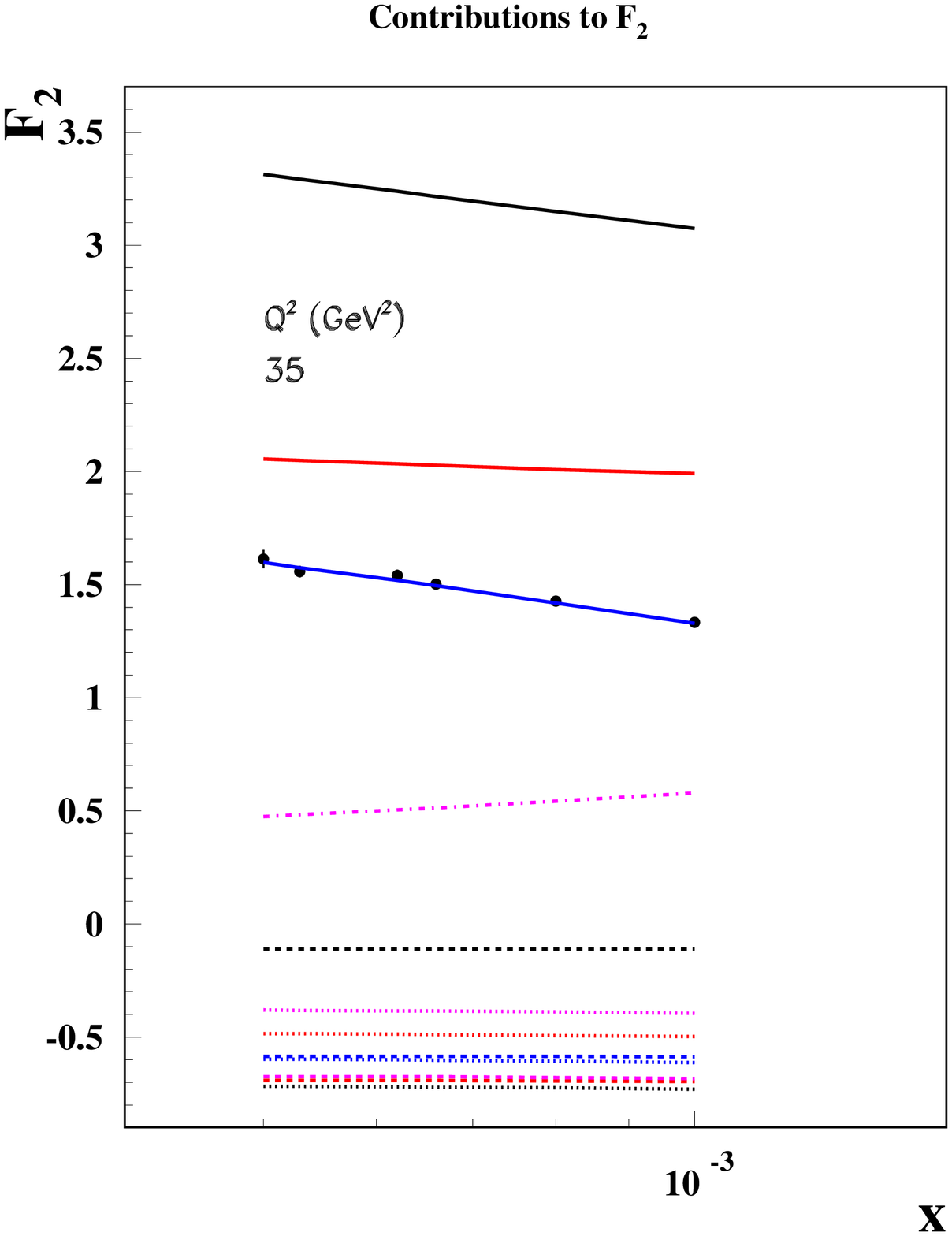}
}
\caption{  Contributions to $F_2$ of individual eigenfunctions. The dots show the measured points at $Q^2 = 6.5$ and 35 GeV$^2$, The full blue line shows the BFKL prediction at these $Q^2$'s, other lines show the contributions of eigenfunctions specified in the legend. With exception of the second eigenfunction and the continuous negative $\omega$ contributions, the contributions of the eigenfunctions are displayed as a sum of two eigenfunctions, (3+4), (5+6), ...  (19+20). }
\label{contr} \end{figure}

 Fig.~\ref{contr}  shows the contributions to $F_2$  from individual eigenfunctions, on the samples
   of results at $Q^2= 6.5$ and 35 GeV$^2$.  The larger dots show the measured points, the full blue lines show the BFKL prediction for $F_2$, similar to  Fig.~\ref{globfit}.  Other lines show the contributions of eigenfunctions specified in the legend, i.e. the terms
\begin{equation}
 F_2^{(n)}(x,Q^2) \ = \ \int_x^1 d\zeta \int \frac{dk}{k} \Phi_{\mathrm{DIS}}(\zeta,Q,k)
    \int \frac{dk^\prime}{k^\prime} \Phi_p(k^\prime)
 \left(\frac{k^\prime \, x}{k}\right)^{-\omega_n} k^2     f^*_{\omega_n}(k^\prime) f_{\omega_n}(k)  
 \label{f2-contr}
\end{equation}
  With exception of the contributions of the second  and of the continuous negative $\omega$ terms, the contributions of other eigenfunctions are displayed as a sum of two eigenfunctions, (3+4), (5+6), ...  (19+20),  to simplify the picture.  The black full line shows the contribution of the second, leading  eigenfunction, which is  substantially larger than $F_2$. 
  
  The contribution of the second eigenfunction, together with the contribution (3+4) and the 
 contribution from the continuum with  negative $\omega$, is  positive.
The contributions of the eigenfunctions 5 to 20 are all negative. The negative contributions correct the positive one to reproduce precisely the measured $F_2$. In this way  the effective slope is also changed; the contribution of the dominating, second term, which has $\omega_2 = 0.144$, is modified to $\lambda = 0.176$ at    $Q^2= 6.5$  GeV$^2$ and  $\lambda = 0.265$ at    $Q^2= 35$  GeV$^2$, in agreement with data.
 Note that the contributions from the subleading eigenfunctions 
are much larger  at  $Q^2= 35$  GeV$^2$  than at    $Q^2= 6.5$  GeV$^2$ due to the increased overlap with the DIS impact factor.  
Note also that the variation of the non-perturbative phases leads to a slower convergence of  the sub-leading terms than in the case of a constant $\eta$, studied in ref.\cite{KLR16}. This is expected because the contribution of the
 subleading terms has to be large enough  to substantially correct  the leading terms in order to reproduce  the data. 
 Nevertheless, we see from Fig.~\ref{contr}   that the contributions of eigenfunctions with $n > 16$  start to approach zero., i.e. show convergence.

Summarising we  can confirm that an excellent description of data is achieved by a fine tune of the non-perturbative phases $\eta_n$. This phase tune is a result of a simple  $\omega -n$ relation which is  well motivated in BFKL and is  determined
  by only two or three parameters. 

\subsection{Decoupling of the first eigenfunction and its consequences}

The decoupling or near decoupling of the first eigenfunction is an unexpected and puzzling feature of this investigation. The decoupling is not connected to a particular value of the proton impact factor or to its form.  The fits of Tables 1 and 2 together with the example of Fig.~\ref{eigenblow} show that when we substantially change the value of the proton impact factor, from $b=10$ GeV$^2$ to  $b=20$ GeV$^2$,  the values of the fit parameters are re-tuned such  that the resulting phases, 
although sightly changed, reproduce  the data  very well and again lead to the decoupling of the leading eigenfunction. 
Note, that these re-tunes hardly change the physical properties of the solution, i.e. the position of the poles,
 owing to the interplay between the phases eigenfunctions and the parameters AB(C).  
A similar result is obtained when we choose a completely different impact factor, given by a delta function, $\Phi_p = A\, \delta(k-k_0)$. Although this is not a realistic impact factor, the results are similar; the fit selects  
the phase of the first eigenfunction  such  that  $f_{\omega_1}(k_0)=0$. The other parameters are  re-tuned so that the fit reproduce the data with a $\chi^2 $ value close to 33, as in the fits of Tables 1 and 2.  

From the technical point of view this  decoupling occurs because the position of the critical point of the first eigenfunction is relatively close to the physical region, $k_c(1) \approx 50 $ GeV, whereas the critical point of the subsequent eigenfunctions  is far away from it,  $k_c(2) \approx 3,3 $ TeV,  $k_c(3) \approx 270 $ TeV, $k_c(4) \approx 20 000 $ TeV, etc.
Therefore, the first eigenfunction varies more quickly near $k_0$ than the subsequent ones, so that a very small change in the
 phase,  $\eta_1$,  leads to a large change of the first contribution.   

We have also checked
 that the results do not depend on  the number of eigenfunction used in the fit, provided this number exceeds 10. 
 In Table 3 we show  the results of fits made with the first 20, 16, 12 and 10 eigenfunctions. All the
fits were made with the ABC relation and in all cases the fit has chosen  a  phase which  decouples the first eigenfunction  from the proton.  
\begin{table}[h]
\begin{center}
 \begin{tabular}{||c|c|c|c|c|c|c|c|||} \hline 
  $N_{ef}$  & 20 & 16 & 12 & 10   \\ \hline
   $A$        &  0.51768 &  0.47904 & 0.44987  &  0.42753  \\  \hline
   $B$ &   1.58209   &  1.32672 & 1.16597 & 1.95858  \\  \hline
   $C$ &   0.000037  &  0.002092 & 0.00431 & 0.00586  \\  \hline
    $\eta_{neg}$ &  -0.0895  &   -0.0723  & -0.0738 & -0.0770  \\ \hline
    $\chi^2$ &  33.4  &  34.0  & 34.4  & 34.7\\ \hline 
 \hline \end{tabular}
\caption{ Results of the ABC-Fit performed with different number of eigenfunctions, $N_{ef}$. All fits were using the same 51 data points, with $x<0.001$ and $Q^2> 6$ GeV$^2$. The value of the proton impact factor was $b=10$ GeV$^2$.}
\end{center}   \end{table}

We conclude therefore that the decoupling or near decoupling of the first eigenfunction is a genuine property of this analysis, independent of the choice of the proton impact factor or the number of eigenfunctions used in the fit.

It is obvious, that this decoupling  can only happen because the leading eigenfunction
 makes a transition from the negative to positive values in a region close to the starting point $k_0$. Such a transition is an indication that the first eigenfunction,  as chosen by the fit, cannot be a  wave function of a ground state because the ground state has to be completely positive, see  Appendix B. 
Therefore, the decoupling of the first eigenfunction should be interpreted as an indication that there exist an additional ground state, corresponding to $n=0$. 
 
 Our computation gives us  some hints about the properties of such a state. From the values of the  turning points, $t_c(n)$, which grows  almost  linearly with $n$, Fig.~\ref{tc}, we can estimate the $k_c$  value of the ground state,  $n=0$, as being around 700  MeV~\footnote{ Taking as example the b=10 GeV$^{-2}$ fit, the $t_c$ values of the first five eigenstates are $t_c(1) = 10.332$,  $t_c(2) = 18.838$,  $t_c(3) = 27.429$,  $t_c(4) = 36,306$,  which correspond to the characteristic momenta of  $k_c(1) \approx 50$ GeV,   $k_c(2) \approx 3,3 $ TeV,  $k_c(3) \approx 260 $ TeV, $k_c(4) \approx 21000 $ TeV. Taking as $\Delta t= t_c(2) -t_c(1) \approx 8.5$ we obtain from $t_c(0) = t_c(1) - \Delta t $ a value  $k_c \approx 700$ MeV. Other values of $k_c(0)$ can be obtained by noting that the increment $\Delta t$ varies slightly with increasing $n$. },  just below our starting value of $k_0 = 1$ GeV.  Such a state would have a high intercept, $\omega_0 \approx 0.3$, and would not have any oscillations above $k_0 $, it would  just  decay exponentially with increasing $\ln(k)$. 
 
\begin{figure}[htbp]
\centerline{
\includegraphics[width=10.0cm,angle=0]{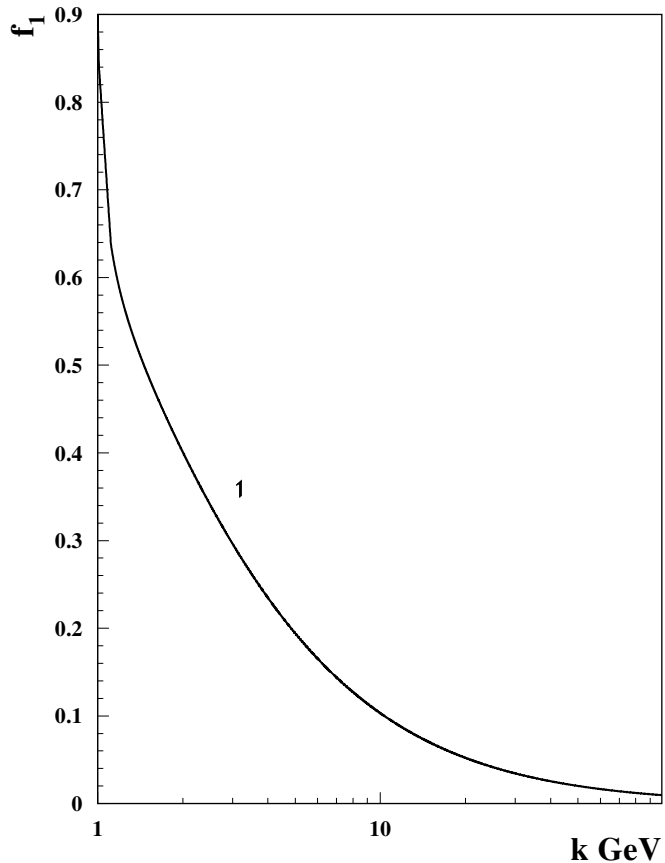}
}
\caption{  Momentum distribution of a state  similar to the real ground state, with $k_c = 1.05 $ GeV, $\omega =0.37$ and $\eta = -2.35$. }
\label{ground} \end{figure}

 As example of such a state we show in Fig.\ref{ground} the momentum distribution of a state which could be similar to the real ground state and which exists in our computation.\footnote{the present numerical setup of the computation does not allow to modify $k_0$ easily. }  It has
 $k_c = 1.05 $~GeV, $\omega =0.37$ and $\eta = -2.35$.

 Indeed, the $k_c$ value of the additional ground state,  of around 700 MeV, lays right in the middle of the saturation region~\cite{GLR,MLV,Muel-Triant,GBW,MSM,BGK,KT,KMW}, where multiple pomeron exchanges should dominate~\cite{Kovch}. In our approach, these exchanges would almost entirely involve the interaction of the low $k_c$ ground state with itself, since  its  size  is much larger than the size of  higher eigenfunctions and the eigenfunctions are orthogonal to each other.  This will lead to  unitarisation  (saturation) corrections which would   substantially affect the properties of the ground state.  The momentum distribution will be shifted 
  towards the lower $k$ values  and therefore its overlap with the photon impact factor should 
diminish  quickly  with increasing $Q^2$. In addition, the saturation correction will damp  the effective exponent of the first eigenfunction, $\omega \approx 0.3$,  to  a value which is compatible with the non-perturbative pomeron state, $\lambda \approx 0.1$.\footnote{ this is known from e.g.  the analysis  of HERA data in terms of the  Golec-Wuesthoff or BGK model~\cite{GBW,BGK}}

  It was already pointed out by Gribov~\cite{Grib}, in the framework of the reggeon calculus, that the soft pomeron could be given by the  renormalised, bare pomeron. The renormalisation procedure should take into account the corrections due to  multiple interactions.  This is somewhat similar to the picture  emerging from our analysis. Of course, the soft pomeron discussed by Gribov, was essentially a non-perturbative state, determined mostly by nuclear forces.\footnote{one of us (HK) would like to thank Al Mueller for an illuminating  discussion on this subject.} In our case, the bare ground state is, however, a perturbative state and its multiple interaction  are also of perturbative origin. Its properties are thought determined, to large extent, by the non-perturbative, nuclear forces which enter into our analysis through the choice of the non-perturbative  
 phase $\eta$. 
  


\subsection{$Q^2$ dependence}
In Table 4 we show the AB-Fit results for different $Q^2$ regions, $Q^2>4, \ 6$ and 9 GeV$^2$, for $b=10$ GeV$^{-2}$ as an example. The fits with $b= 20 $ GeV$^2$ and/or  the $ABC$ fits show very similar results.  \begin{table}[h]
\begin{center}
 \begin{tabular}{||c|c|c|c|c|c|c|c|||} \hline 
  $Q^2$ cut (GeV$^{2}$)  & 4 & 6 & 9   \\ \hline
   $A$               & 0.51852  &  0.51844  &  0.51818    \\  \hline
   $B$               & 1.58847  &   1.58697   &  1.58356    \\  \hline
   $\eta_{neg}$ &  -0.0911  &  -0.0911  &   -0.0911  \\ \hline
   $N_p$           &  59 &  51  &   37  \\ \hline
    $\chi^2$       & 68.5  &  33.9  & 17.4  \\ \hline 
    $\chi^2/N_{df}$     & 1.25  &  0.72  & 0,52  \\ \hline 
 \hline \end{tabular}
\caption{ Results of the AB-Fit  with $x<0.001$ and $b=10$ GeV$^{-2}$. }
\end{center}   \end{table}
 The fit   with  $Q^2>4$ GeV$^2$ of  Table 4  has a substantially lower quality than the one with $Q^2>6$ GeV$^2$. 
 Also the fit with  $Q^2>6$ GeV$^2$ is significantly worse than the  $Q^2>9$ GeV$^2$ one. 
 Therefore, it is possible  that the worsening of the fit quality with decreasing $Q^2$ cut is  due to the presence of the hypothetical ground state discussed above.  
 

\subsection{Extrapolation to very low $x$}
\begin{figure}[htbp]
\centerline{
\includegraphics[width=14.0cm,angle=0]{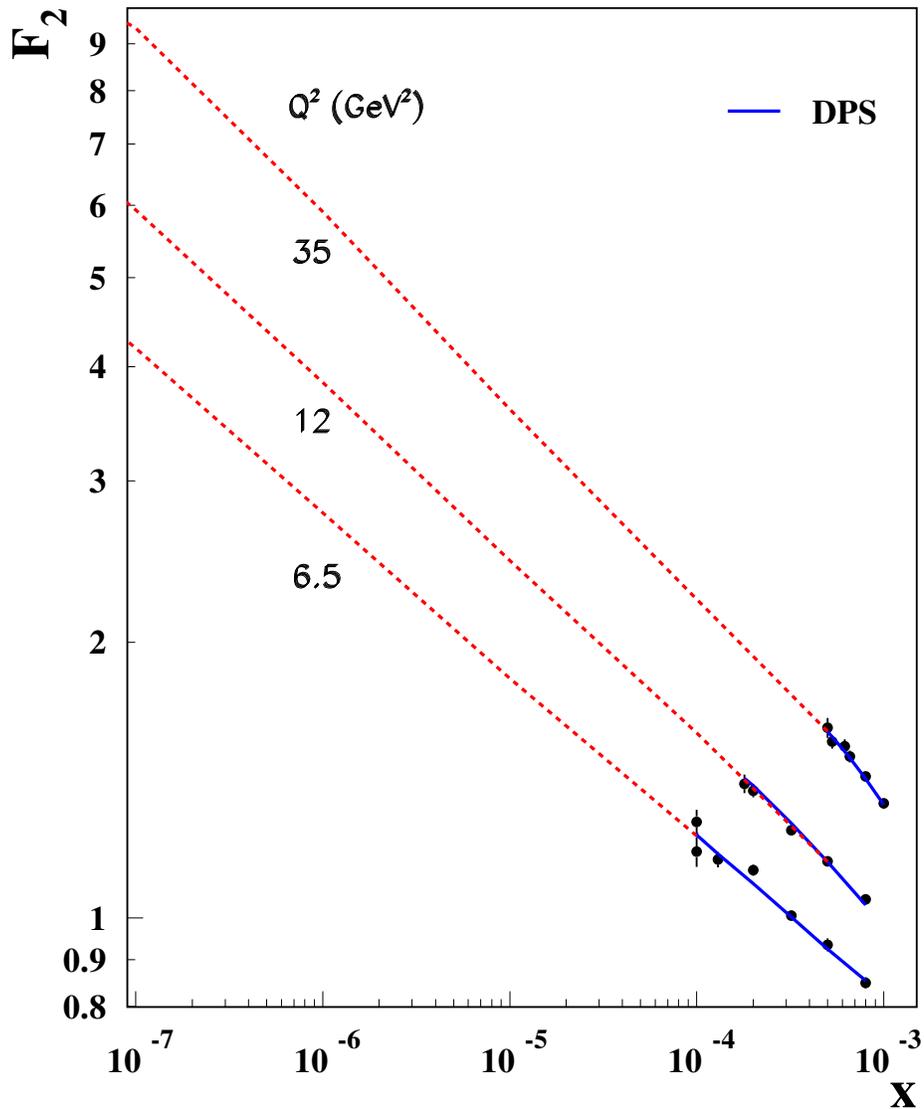}
}
\caption{ Extrapolation of the AB fit results to very low $x$   }
\label{globext} \end{figure}

In Fig.~\ref{globext} we show the extrapolation of the $AB$ fit to very low $x$ values, which can be possibly achieved in some  future $ep$ collider like VHEeP or LHeC. We see that at very large energies the increase of $F_2$  shows similar slopes at different $Q^2$ values, unlike at HERA. This is due to the dominance of the leading trajectory at very low $x$ values. 
 
 \newpage
\section{Conclusions and Outlook}
\label{sec10}
 We have shown here that there exists an infrared 
  boundary condition, which  leads to a precise description of  HERA $F_2$ data, for $x<0.001$. We formulated it in terms of 
 a  relation between the eigenvalues, $\omega_n$, and the  eigenfunction number, $n$.
 It has  a simple form  $\omega_n = A/(B+n)$ or $\omega_n = A/(B+n) +C$,  called here $AB$ or $ABC$ relations respectively. Both relations are  well motivated in BFKL,  for larger $n$. The $\omega - n$ relation determines, within the BFKL Green Function solution, the values of the phases of the eigenfunctions,  $\eta_n$, close to the non-perturbative region, at  small $k \sim \Lambda_{\rm QCD}$.  The fits using both relations give an excellent description of data with
 similar $\chi^2$ values. 
 
 The fits  lead to the unexpected result that the first eigenfunction  decouples or nearly decouples. This means that  the overlap of the first eigenfunction with the proton  impact factor is very small or even zero, due to the fact that the first eigenfunction has a  transition region from negative to positive values, i.e. a node. Therefore,  the first eigenfunction chosen by the fit cannot be a ground state. This suggests, as a consequence,  the existence of a multiply interacting  ground  state which may have properties of the soft pomeron.    The contributions of such a state would be rapidly attenuated as $Q^2$ increases. However,  at low $Q^2$, it should dominate the $F_2$  and  diffractive processes.  A particularly good place to study its effects should be the exclusive diffractive vector meson production,  $\rho$, $\phi$ and $J/\psi$, because in this reaction the value of the Regge slope, $\alpha'$, is also measured.    We may try to learn more about it in our forthcoming paper by focusing the investigation on the region closer to $\Lambda_{QCD}$, by varying $k_0$ and, last but not least, using the complete information  concerning the errors of HERA data~\cite{HERAFitter}.

 The present  BFKL fits to HERA data predict  that   in the very low $x$ region, $x << 10^{-4}$, $Q^2>6$ GeV$^2$,  $F_2$ should grow with a slope $\lambda $ which is close to the eigenvalue of the second eigenfunction and which is $Q^2$ independent. This prediction is possible because there is no interference between the ground state and the second eigenfunction, since they are orthogonal to each other and have very different support. The $\omega $ value of the second eigenfunction  could be easily measured on some future $ep$ collider, such as VHEeP~\cite{Cald-Wing} or LHeC~\cite{LHeC}.

 Finally, let us note that the AB(C) fits, should be affected by supersymmetry or other physics beyond the standard model (BSM),
   as  discussed in our previous papers~\cite{KLRW,KLR}.  This is because
 (at least at LO)  the constant $A$ is proportional to the beta function, which changes its value drastically once the threshold for the  production
 of gluimos or other BSM particles  is crossed.  
 The decoupling of the leading eigenfunctions
     makes the analysis of BSM physics simpler, especially on the future VHeP or LHeC colliders. This is  because the ground state is now very well constrained, the value of the $\omega_2$ can be directly measured, and the values of the higher intercepts, $\omega_{n > 2}$, can be parametrised reliably.  
 

\newpage
\section{Appendix A}
We rephrase here the original derivation of the BFKL resummation given in ref.~\cite{salam}.
It is convenient to write
 $$\chi_0(\nu) \ \equiv \chi(0,\nu) $$
where
  \beq \chi(a,\nu) \ \equiv \ 2\Psi(1) - \Psi\left(\frac{1}{2}+a+i\nu\right) 
 - \Psi\left(\frac{1}{2}+a-i\nu\right)  \eeq
and
 \beq \dot{\chi}(a,\nu) \equiv \ \frac{d}{da} \chi(a,\nu)
 \ = \ - \Psi^\prime\left(\frac{1}{2}+a+i\nu\right) 
 - \Psi^\prime\left(\frac{1}{2}+a-i\nu\right)  \eeq
If $a$ is small then up to order $a$ we have
 \beq \chi(a,\nu) \ = \ \chi(0,\nu) + a \dot{\chi}(0,\nu)
 + {\cal O}\left( a^2\right), \label{expand} \eeq

We may write $\chi_1(\nu)$ (defining a quantity $\chi_1^{reg}(\nu)$ ) as
\beq \chi_1(\nu) \ \equiv \ -A \chi(0,\nu) + B \dot{\chi}(0,\nu)
  + \frac{1}{2} \chi(0,\nu) \dot{\chi}(0,\nu) + \chi_1^{reg}(\nu) \eeq

By a suitable choice of the constants $A$ and $B$, we can arrange for
$\chi_1^{reg}(\nu)$ to be free of singularities as 
$\nu \, \to  \pm \frac{i}{2} $.
 
In this limit we have
 \beq \chi(0,\nu) \stackrel{\nu\to\pm i/2}{\longrightarrow}
 \frac{1}{\left(\frac{1}{2}\pm i \nu\right)} 
   +{\cal O} \left(\frac{1}{2}\pm i \nu\right) \eeq

and

 \beq \dot{\chi}(0,\nu) \stackrel{\nu\to\pm i/2}{\longrightarrow}
 -\frac{1}{\left(\frac{1}{2}\pm i \nu\right)^2}  +\frac{\pi^2}{3}
   + {\cal O} \left(\frac{1}{2}\pm i \nu\right)  \eeq

So that
 \beq \chi_1(\nu) \stackrel{\nu\to\pm i/2}{\longrightarrow}
 -\frac{1}{2\left(\frac{1}{2}\pm i \nu\right)^3} 
 -\frac{B}{\left(\frac{1}{2}\pm i \nu\right)^2} 
 -  \left(A+\frac{\pi^2}{6} \right)
\frac{1}{\left(\frac{1}{2}\pm i \nu\right)} \eeq

Therefore the constants $A$ and $B$ are selected to match the single and double
poles respectively of the function $\chi_1(\nu)$ and in that way
 $\chi_1^{reg}$ is free from such singularities.

 in ref,\cite{salam} it is pointed out that the correction due to $\chi_1^{reg}$ is genuinely
negligible and the entire large correction to the characteristic function
come from the terms which are singular as $\nu \to \pm i/2$.

Now let us consider another function $\tilde{\omega}(\nu)$
which is defined as the solution to the transcendental (implicit) 
 equation
\beq \tilde{\omega}(\nu) \ \equiv 
 \alphabar(1-\alphabar A)  \chi\left(\frac{\tilde{\omega}}{2}+\alphabar B,\nu \right)
  + \alphabar^2 \chi_1^{reg}(\nu) \eeq

Solving to leading order in $\alphabar$ we have
 \beq \tilde{\omega} \ = \ \alphabar  \chi(0,\nu) + {\cal O}\left(\alphabar^2)\right) .\eeq
Expanding  $\tilde{\omega}(\nu)$ up to order $\alphabar^2$, and using
 (\ref{expand}) we obtain
\begin{eqnarray}  \tilde{\omega}(\nu)  & = & \alphabar(\chi(0,\nu)
\ + \ \alphabar^2\left[-A \chi(0,\nu) + B \dot{\chi}(0,\nu)
  + \frac{1}{2} \chi(0,\nu) \dot{\chi}(0,\nu) + \chi_1^{reg}(\nu)\right]
 + {\cal O}\left(\alphabar^3\right)
\nonumber \\ &=& 
 \alphabar(\chi_0(\nu) + \alphabar^2 \chi_1(\nu)
 + {\cal O}\left(\alphabar^3\right)  
\label{eq11}  \end{eqnarray}

Thus we see that up to order $\alphabar^2$, the quantities $\omega(\nu)$
and $\tilde{\omega}(\nu)$ are identical so that up to that accuracy
 we may replace the usual perturbative expression given in (\ref{pert13})
 by $\tilde{\omega}(\nu)$.

On the other hand, the quantity $\tilde{\omega}(\nu)$ {\it does not
contain any  singularities} as $\nu \, \to \, \pm \frac{i}{2} $.
The  singularities we see in eq(\ref{eq11}) are {\it only present 
 as a result of an expansion}. They are therefore an artifact of this expansion
and are not present for the entire function.
Since it is these singular terms that give rise to the large NLO
corrections found in $\chi_1(\nu)$ we may consider the quantity
 $\tilde{\omega}(\nu)$ to be the expression in which all of these
 large corrections have been resummed. 

For the case of the third order pole, this has been established exactly, since we
know what the origin of the triple pole is. In rev\cite{salam} it is  explained
  that this arises
 from a mismatch between the ``rapidity'', $Y$, of the forward gluon-gluon
scattering amplitude used in the BFKL approach
 $$ Y \ = \ \ln\left(\frac{s}{k k^\prime}\right) $$

For the resummation of the double and  single poles, this is not
known uniquely and there are an infinit enumber of possible  resummation schemes,
 of which  one is described here, and three oithers are discusswd in ref.\cite{salam}.
 All these reummation schemes  have in common the fact  that they  resum
 all the collinear singularities (i.e. all poles as
 $\nu \, \to \, \pm \frac{i}{2}$ and they are all
 equivalent to the ordinary pertubative expansion for $\omega$
 up to order $\alphabar^2$. They, of course, differ, in the
terms proportional to $\alphabar^3$ and higher - but we have no
 reason to select one of these schemes above another in the absence
of the NNLO calculation of the characteristic function. Scheme 3, which is
the scheme considered here is the most convenient for our purposes.

\newpage
\section{Appendix B}


\bigskip

\centerline{\large \bf Absence of nodes in the wave function of a ground state}


\vspace{7mm}
 One can define  the kinetic energy,  $\widetilde{T} [\psi ]$,  as  
\beq
\widetilde{T}[\psi ]=-\frac{1}{2m}\,\int _{-\infty}^{\infty}\psi (x)\,\psi''(x)dx .
\label{kin1}
\eeq
Integrating by parts we obtain  
\beq
T[\psi ]= \frac{1}{2m}\,\int _{-\infty}^{\infty}((\psi'(x))^2dx ,
\label{kin2}
\eeq
provided the wave function $\psi(x)$ is continuous and has continuous first derivatives. (The transition from (\ref{kin1})  to  (\ref{kin2}) is not valid for  the continuous wave functions which do not have a fully continuous first derivatives, like e.g. $\psi (x)\sim  \alpha |x|$ 
or $\psi (x)\sim \exp (-\alpha |x|)$.) In the following, we prefer to use for kinetic energy the expression (\ref{kin2}) since,  in contrast to  (\ref{kin1}), it is always positive.  

Let us first consider the case of  the one-dimensional Schr\"{o}dinger equation and   define  the total energy   as a functional 
\beq
E[\psi]=T[\psi]+V[\psi]=\frac{1}{2m}\,\int _{-\infty}^{\infty}(\psi'(x))^2dx+\int _{-\infty}^{\infty}(\psi (x))^2V(x)\,dx \,.
\eeq
In the case of a ground state of  energy $E_0$, the functional  $E[\psi ]$  takes the minimal value calculated on all possible normalized wave functions
\beq
E_0=\min _{\psi} \,\frac{E(\psi)}{||\psi||^2}\,,\,\,||\psi||^2\equiv \int _{-\infty}^{\infty}(\psi (x))^2\,dx\,.
\eeq

Let us assume, that the $\psi$-function changes its sign, for example, $\psi (x)|_{x\rightarrow 0} \sim x$, and prove, that there is a positive function $\chi (x)$ with $\chi (0)\ne 0$, which has a smaller energy $E$.
 It would mean,
that the wave function $\psi$ with a node at $x=0$ cannot be the wave function of the ground state.

We choose   the trial wave function $\chi (x)$ in the form
\beq
\chi (x)|_{_{|x|>\epsilon}}\equiv |\psi (x)| \,,\,\,\chi (x)|_{_{|x|<\epsilon}}\equiv \frac{|\psi '(0)|}{2|\epsilon|}\,\left(x^2+\epsilon^2\right)\,,\,\,
\psi (x)|_{_{|x|<\epsilon}}\approx |\psi' (0)|\,x\,,
\eeq
where $\epsilon \rightarrow 0$. Note, that $\chi (x)$ is a continuous function having also continuous derivatives at $x=\pm \epsilon$. One can neglect small corrections $\sim \epsilon ^3$  to the normalisation integral and to the potential energy $V(\chi)$. The main contribution to $\delta E(\chi )$ is obtained from the kinetic energy 
\beq
\delta E=T(\chi )-T(\psi )=\frac{1}{2m}\int _{-\epsilon}^{\epsilon}(\psi'(0))^2\left(x^2/\epsilon^2-1\right)dx=
-\frac{2(\psi'(0))^2\epsilon}{3m} \,.
\eeq
Because $\delta E<0$ we conclude that, in case of the Schr\"{o}dinger equation,  the ground state wave function cannot have nodes.
$$
$$
Let us turn  now to the BFKL equation with the running coupling constant. In the leading logarithmic approximation we have
\beq
-\omega\,f=H_{BFKL}\,f\,,\eeq
with 
\beq
H_{BFKL}=\sqrt{\alpha_s(t)}\,\left(\Psi \left(\frac{1}{2}+i\nu \right)+\Psi \left(\frac{1}{2}-i\nu \right)-2\Psi \left(1 \right)\right)\,\sqrt{\alpha_s(t)}\, ,
\eeq
where
\beq
\alpha_s(t) =\frac{1}{\bar{\beta}_0t}\,,\,\,t= \ln \frac{|k_\perp|^2}{\Lambda^2_{QCD}}\,.
\eeq
Here $E=-\omega$ plays the role of the total energy in the Schr\"{o}dinger equation.
The operator $\nu$ denotes the momentum  canonically conjugated to the coordinate $t$,
\beq
[\nu , t]=i\,.
\eeq
As usual in QCD, one can use the perturbative hamiltonian $H$  for large $t>t_0>0$ only. For $t<t_0$ it should be substituted by an hermitian non-perturbative
hamiltonian $\widetilde{H}$ and the corresponding wave functions and their derivatives are matched at $t=t_0$.

We prove now that the ground state wave function $f_0$, with energy $E_0$,  cannot have a node at $t=t_1>t_0$.
  For this purpose, as  in the above case of the usual quantum mechanics, we use a simple trial function $\chi (t)$, which is different from $|f(t)|$ (with $f(t_1)=0$)  only in the small region $\sim \epsilon$ around $t=t_1$
\beq
\chi (t)|_{_{|t-t_1|<\epsilon}}\equiv \frac{|f '(t_1)|}{2|\epsilon|}\,\left((t-t_1)^2+\epsilon^2\right)\,,\,\,f(t)|_{_{|t-t_1|<\epsilon}}\approx |f'(t_1)| (t-t_1)\,,\,\,\epsilon \ll 1\,.
\eeq
Note, that for the BFKL hamiltonian, which has a non-linear dependence from $\nu^2$, it would be natural to introduce a trial function $\chi$ with  continuous higher derivatives in the points $t-t_1=\pm \epsilon$. But in the correction to the total energy, expressed in terms of the functional
\beq
E=\int dt \,f(t)\,H\,f (t)\,,\,\,||f||=1
\eeq
with
the substitution $f(t)\rightarrow |f(t)|\rightarrow  \chi(t)$,
the contribution from the region $|t-t_1|>\epsilon$ will ramain unchanged. In the region $|t-t_1|<\epsilon$, the higher derivatives of the BFKL hamiltonian $H$, acting on the simple polynomial functions $\chi (t)$ and $f (t)$, should be neglected.  Note that this corresponds to the diffusion approximation, because only terms proportional to $\nu^2$, in the expansion of the hamiltonian $H$, should be taken into account.

As above,  corrections to the normalisation condition and to the running coupling factors $\sqrt{\alpha_s(t)}$ are small. Thus, the main
correction to the total energy  of the trial function can be written as 
\beq
\delta E=\int _{t_1-\epsilon}^{t_1+\epsilon}\alpha_s (t_1)\,14 \zeta (3)\,(\chi^{\prime \,2}(t)-f^{\prime \,2}(t))dt=
-14 \zeta (3)\,\alpha _s(t_1)\,f^{\prime \,2}(t_1)\frac{4\epsilon}{3} \,  ,
\eeq
  when $\epsilon \rightarrow 0$.
Because this correction is negative we conclude that the ground state wave function for the BFKL pomeron cannot have nodes.

\newpage
																									
\section*{Acknowlegments}
We are grateful to Jochen Bartels,  Al Mueller, Agustin Sabio-Vera, Anna Stasto and Gia Dvali for useful
conversations.
One of us (LL) would like to thank  the State University of St. Petersburg for the grant
SPSU 11.38.223.2015  and the grant  
 RFBI 16-02-01143 for support.
One of us (DAR) wishes to thank the Leverhulme Trust for an
Emeritus Fellowship.

\end{document}